\documentclass[a4paper]{aa}
\ifx\pdfoutput\undefined
\usepackage[dvips]{graphicx}
\else
\usepackage{graphicx}
\fi
\usepackage{epstopdf}
\usepackage[latin1]{inputenc}
\usepackage{amsmath}
\usepackage{amsfonts}
\usepackage{amssymb}
\usepackage{graphicx}
\usepackage{txfonts}
\usepackage{natbib}
\usepackage{multirow}
\bibpunct{(}{)}{;}{a}{}{,} % to follow the A&A style

\begin{document}
\DeclareGraphicsExtensions{.pdf,.gif,.jpg}
 \bibliographystyle{aa}

\title{A weak lensing analysis of the Abell 383 cluster.
\thanks{Based  on: data collected at Subaru Telescope  (University of Tokyo) and obtained from the SMOKA, which is operated by the Astronomy Data Center, 
National Astronomical Observatory of Japan;  observations obtained with MegaPrime/MegaCam, a joint project 
of CFHT and CEA/DAPNIA, at the Canada-France-Hawaii Telescope (CFHT) which is 
operated by the National Research Council (NRC) of Canada, the Institute 
National des Sciences de l'Univers of the Centre National de la Recherche 
Scientifique of France, and the University of Hawaii. This work is based in part on data products produced at TERAPIX and the Canadian Astronomy Data Centre as part of the Canada-France-Hawaii Telescope Legacy Survey, a collaborative project of NRC and CNRS.}
}
\author{Zhuoyi Huang\inst{1} \and Mario Radovich \inst{2} \and Aniello Grado\inst{1} \and Emanuella Puddu \inst{1} \and Anna Romano\inst{3}   \and \\Luca Limatola\inst{1} \and Liping Fu\inst{4,1}}

\institute{
INAF - Osservatorio Astronomico di Capodimonte, via Moiariello 16, I-80131, Napoli, Italy
\and INAF - Osservatorio Astronomico di Padova, vicolo dell'Osservatorio 5, I-35122, Padova, Italy
\and INAF - Osservatorio Astronomico di Roma, Monte Porzio, I-00185, Roma, Italy
\and Key Lab for Astrophysics, Shanghai Normal University, 100 Guilin Road, 200234, Shanghai, China }
\date{received; accepted}

 \abstract
   {} %leave it empty if necessary
   {In this paper we use deep CFHT and SUBARU $uBVRIz$ archival images of the Abell 383 cluster (z=0.187) to estimate its mass by weak lensing.}
   {To this end, we first use simulated images to check the accuracy provided by our KSB
pipeline. Such simulations include both the STEP 1 and 2
simulations, and more realistic simulations of the distortion of
galaxy shapes by a cluster with a Navarro-Frenk-White (NFW) profile. From such simulations
we estimate the effect of noise on shear measurement and derive the
correction terms.  The $R-$band image is  used to derive the mass by
fitting the observed tangential shear profile with a
NFW mass  profile. Photometric redshifts are
computed from  the $uBVRIz$ catalogs. Different methods for the
foreground/background galaxy selection are implemented, namely
selection by magnitude, color and photometric redshifts, and results
are compared. In particular, we developed a semi-automatic algorithm
to select the foreground galaxies in the color-color diagram, based
on observed colors. }
   {Using color selection or photometric redshifts improves the
correction of dilution from foreground galaxies: this  leads to higher
signals in the inner parts of the cluster. We obtain a cluster mass
$M_{\rm vir} = 7.5^{+2.7}_{-1.9} \times 10^{14}$ $M_\odot$: such
value is $\sim$ 20\% higher than previous estimates, and is more
consistent the mass expected from X--ray data. The $R$-band luminosity
function of the cluster is finally computed, giving a total luminosity
$L_{\rm tot}=(2.14\pm0.5)$ $\times$ $10^{12}~L_\odot$
and a mass to luminosity ratio $M/L \sim 300 M_\odot/L_\odot$.}
   {}


\maketitle

\section{Introduction}
\label{sec:intro}

Weak gravitational lensing is a unique technique that allows to probe the
distribution of dark matter in the Universe.
It measures the very small distortions in the shapes of faint background
galaxies, due to foreground mass structures.
This technique requires a very accurate measurement of the shape parameters as
well as the removal of the systematic effects affecting them.
In addition, galaxies to be used in the weak lensing analysis must
be carefully selected so that they do not include a significant fraction of unlensed sources,
with redshift smaller than that of the lens. This would introduce a dilution
and therefore an underestimated signal, in particular towards the cluster center
 (see \citealt{Broadh05}): such effect may be the reason of the
under-prediction by weak lensing of the observed Einstein radius, by a factor of $\sim2.5$ \citep{smith01, Bardeau05}.
The optimal case would happen if  photometric redshifts were available. Even if for weak lensing high accuracy in their estimate are not 
required  for individual galaxies, on average we need at least $\sigma_z/(1+z) < 0.1$: 
this implies having observations in several bands, spanning a good wavelength range.
If few bands are available, an uncontaminated background sample can be  obtained
by selecting only galaxies redder than the cluster red sequence \citep{Broadh05}. However, such
method often does not allow to get a number density of background  sources high enough to allow an
accurate weak lensing measure. Including galaxies bluer than the red sequence  \citep{Okabe10}
requires a careful selection of the color offset, as bluer galaxies can be still contaminated by late--types members
of the cluster.
Finally, if  more than two bands are available, \citet{Medez10} discussed how
 to identify  cluster members  and the foreground population as overdensities in the color-color space.

In this paper we exploit deep $uBVRIz$ images of the cluster Abell 383,
taken with the MEGACAM and SUPRIME camera mounted on the 3.6m CFHT and  8m SUBARU telescopes respectively,  and publicly available. 
The mass of the cluster is derived by weak lensing, and values obtained by different selection methods are compared.
The properties of the cluster are reviewed in  Sect.\ref{sec:a383}. Data reduction is discussed in Sect.\ref{sec:data}.
In Sect.~\ref{sec:ksbex} we describe the algorithm used for the shape
measurement, and some improvements for the removal of  biases. The accuracy in the mass estimate is derived by comparison with
simulations.
In Sect.~\ref{sec:results}, we first summarize the different methods for the selection of the background galaxies
from which the lensing  signal is measured. Such methods are applied to the case of Abell 383, and  the
masses derived in such way are then compared. Finally, in Sect.~\ref{sec:discussion} we compare the mass derived in this paper
with  literature values, both by X--rays and weak lensing; a comparison is also done with the mass expected for the $R$--band luminosity, derived from  the luminosity function of Abell 383.

A standard cosmology was adopted in this paper: $\Omega_\Lambda = 0.7$, $\Omega_M = 0.3$, $H_0 = 70$ km s$^{-1}$ Mpc$^{-1}$, giving a scale of 2.92 kpc/arcsec
at the redshift of Abell 383.

% Several methods have been proposed in the last years, which mainly differ in
% how the Point Spread Function (PSF) is removed. Basically they fall in two main groups:
% 1) KSB-like; 2) fitting\citet{step1}.

\section{Abell 383}
\label{sec:a383}

Abell 383 is an apparently well relaxed cluster of galaxies of richness class 2 and of Bautz-Morgan type II-III \citep{Abell89},
located at z = 0.187 \citep{Fetisova93}.
It is dominated by the central cD galaxy, a blue-core emission-line
Bright Cluster Galaxy (BCG), that is aligned with the X-ray peak \citet{smith01}.
Abell 383 is one of the clusters of the XBACs sample (X-ray-Brightest Abell-type
Clusters), observed in the ROSAT All-Sky Survey (RASS; \citealt{Voges92}):
its X-ray luminosity is $8.03\times 10^{44}$ erg sec$^{-1}$ in the 0.1-2.4 keV band
and its X-ray temperature is 7.5 keV \citep{ebeling}. A small core radius, a
steep surface brightness profile and an inverted deprojected temperature profile,
show evidence of the presence of a cooling flow, as supported by the strong
emission lines in the optical spectra of its BCG \citep{Rizza98}.

An extensive study of this cluster was carried out by \citet{smith01}, in
which lensing and X-ray properties were analyzed on deep optical HST images and ROSAT HRI data,
respectively. A complex system of strong lensed features (a giant arc, two radial arcs
in the center and numerous arclets) were identified in its HST images,
some of which are also visible in the deep SUBARU data  used here.

\section{Data retrieval and reduction}
\label{sec:data}

The cluster Abell 383 was observed with the SUPRIME camera  mounted at the 8m SUBARU telescope:
SUPRIME is a ten CCDs mosaic, with a $34 \times 27$ arcmin$^2$ field of view \citep{suprimecam}.
Data are publicly available in the $BVRIz$ filters, with total exposure times of 7800s ($R$), $\sim$ 6000s ($B$, $V$),
3600s ($I$) and 1500s ($z$);
they were retrieved using the SMOKA\footnote{http://smoka.nao.ac.jp/} Science Archive facility.
Data were collected from seven different runs,  amounting to $\sim  55$GB.
Details about the observation nights and exposure times for each band are given in Table~\ref{ObsLog}.
The  data reduction was done using the  VST--Tube
imaging  pipeline, specifically  developed for  the VLT Survey Telescope   \citep[VST,][]{Capaccioli} but
adaptable to other existing or future multi-CCD cameras \citep{Grado10} .

The field of Abell 383 was also observed in the $u^*$ band with the MEGACAM camera attached on the Canada-France Hawaii 
Telescope (CFHT), with a total exposure time of 10541s. The preprocessed images were retrieved from the CADC archive.

\begin{figure}
 \centering
 \includegraphics[width=7 cm,angle=270]{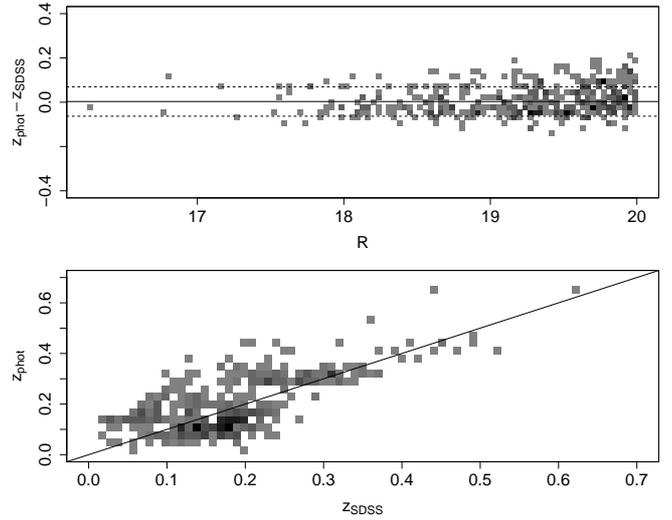}
 % apmass.png: 970x797 pixel, 85dpi, 28.99x23.82 cm, bb=0 0 822 675
 \caption{Density plots comparing the photometric redshifts in the Abell 383 field available from the SDSS, and those here computed from the $uBVRIz$ photometry.}
 \label{fig:sdss}
\end{figure}

\begin{figure}
 \centering
 \includegraphics[width=7 cm,angle=270]{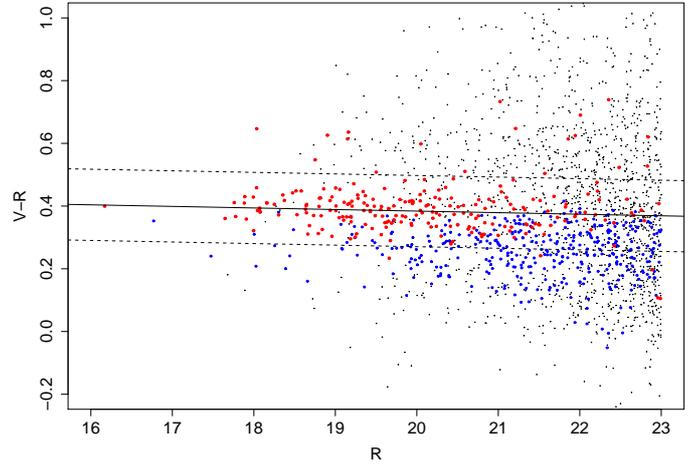}
 % apmass.png: 970x797 pixel, 85dpi, 28.99x23.82 cm, bb=0 0 822 675
 \caption{$V-R$ vs $R$ color plot: red and blue points are galaxies at $z_{\rm phot} = 0.187  \pm$ 0.1 and
classified as early and late-type, respectively; dashed lines shows the $\pm 1 \sigma$ levels.}
 \label{fig:redseq}
\end{figure}

%\subsection{Data reduction}

The basic reduction steps were performed for each frame, namely  overscan correction, flat fielding, correction of the geometric distortion due to the optics, and sky background subtraction.
It is worth  mentioning that to improve the photometric accuracy, a sky superflat was used.

%__________________________________________________ One column table
   \begin{table}
   \begin{center}
      \caption[]{Summary of observations done with the MEGACAM ($u$) and SUPRIME ($BVRIz$) cameras, used in this paper.}
         \label{ObsLog}
         \begin{tabular}{lllll}
            \hline\hline
         %   \noalign{\smallskip}
            Date    & Band & Exp. Time & Total \\
         %   \noalign{\smallskip}
            \hline
         %   \noalign{\smallskip}
           
            23 Dec 2003 & $u^*$ & 8381s \\
            21 Jan 2004 & $u^*$ & 2160s \\
            & & &10541s \\
            09 Sept 2002 & $B$ & 2400s \\
            08 Jan 2008 & $B$ & 2400s\\
            09 Jan 2008 & $B$ & 1200s \\
            & & &6000s\\
            02 Oct 2002 & $V$ & 4320s \\
            01 Oct 2005 & $V$ & 1800s\\
            & & &6120s\\
           12 Nov 2007 & $R_c$ & 2400s\\
            08 Jan 2008 & $R_c$ & 5400s\\
            & & &7800s\\
            02 Oct 2002 & $I_c$ & 3600s\\
            && & 3600s\\
             10 Sept 2002 & $z'$ & 1500s \\
	    &&& 1500s\\

          %  \noalign{\smallskip}
            \hline
            \end{tabular}
         \end{center}

   \end{table}

Geometric distortions were first removed from each exposure, using the \textsc{Scamp} tool \footnote{\textsc{Stuff}, \textsc{Skymaker}, \textsc{SWarp} and \textsc{SExtractor} are part of the Astromatic software developed by E. Bertin,  see www.astromatic.net} and taking the USNO-B1 as the astrometric reference catalog. The internal accuracy provided by \textsc{Scamp}, as measured by positions of the same sources in different exposures, is $\sim 0.05$ arcsec. Different exposures were then stacked
together using  \textsc{SWarp}. The coaddition was done in such a way that all the images had the same scale and size.

%using {\bf lino a tool of EFits library \citep{efits} which combines images using a } robust M-estimator (sigma-clip) instead of the commonly used median. This allowed to gain approximately  {\bf lino 0.5 ->0.25} magnitude in the 50\% completeness limit.\\
The photometric calibration of the $BVRI$ bands was performed  using the standard Stetson fields, observed in the same nights of the data. In the case of the $z$ band, we used a pointing also covered by the Stripe 82 scans in the Sloan Digital Sky Survey (SDSS);
the SDSS photometry of sources identified as point-like was used to derive the zero point in the SUBARU image.

In the case of the MEGACAM-CFHT $u^*$ band images, reduced images and photometric zero points are already available from the CADC public archive, hence only  astrometric calibration and stacking were required.

Table~\ref{MagSeeing} summarizes the photometric properties of the final coadded images (average FWHMs and limiting magnitudes for point-like sources) for each band.
All magnitudes were converted to the AB system; magnitudes of sources classified as  galaxies were corrected for Galactic extinction using the Schlegel maps \citep{schlegel}.

The weak lensing analysis was done on the  $R-$band image.
The masking of reflection haloes and diffraction spikes near bright stars are performed by \textsc{ExAM}, a code developed for this purpose. In short, \textsc{ExAM} takes the \textsc{SExtractor} catalog as input, locates the stellar locus in the size-magnitude diagram (see Sect.~\ref{sec:sgclass}), picks out stars with spike-like features from the isophotal shape analysis, and outputs mask region file that may be  visualized in the \textsc{DS9} software, and finally creates a mask  image in FITS format. The reflection haloes are masked by estimating the background contrast near the bright stars, whose positions are obtained from the USNO--B1. The effective area available after removal of regions masked in such way was 801 arcmin$^2$.
Catalogs for the other bands were extracted using  \textsc{SExtractor} in dual--mode, where the $R-$band image was used as the detection image.

\begin{table}
   \begin{center}
      \caption[]{Photometric properties of the coadded images; FWHM (arcsec) and limiting magnitudes (in the AB system)  were computed for point-like sources. The signal to noise (SNR) is defined as SNR=FLUX\_AUTO/FLUXERR\_AUTO.}
         \label{MagSeeing}
         \begin{tabular}{lccc}
            \hline\hline
    %        \noalign{\smallskip}
            Band    & FWHM & mag (SNR=5) & mag (SNR=10)  \\
     %       \noalign{\smallskip}
            \hline
      %      \noalign{\smallskip}
	    $u^*$ & 1.3  & 26.1  & 25.3 \\
            $B$ &  0.99 & 27.0 & 26.2\\
            $V$ &  0.95 & 26.6 & 25.7\\
            $R_c$ &  0.82 & 27.1 & 26.1\\
            $I_c$ &  0.97 & 25.0 & 24.2\\
            $z'$ &  0.74 & 24.7& 23.9\\
       %     \noalign{\smallskip}
            \hline
            \end{tabular}
         \end{center}
   \end{table}

\begin{figure}
 \centering
 \includegraphics[width=7 cm,angle=270]{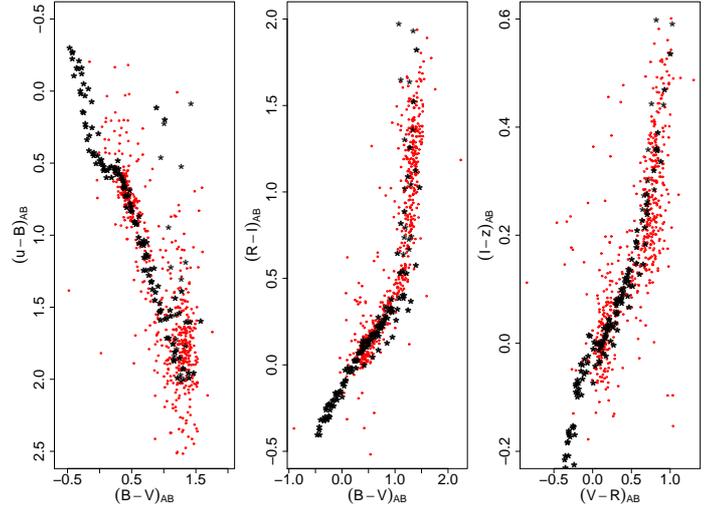}
\caption{Observed (red dots) and model colors (black dots) for stars, after the offsets given in the text were applied. Model colors
were derived convolving the Pickles' library of stellar spectra with the filter transmission curves.}
 \label{fig:stars}
\end{figure}

Photometric redshifts were computed from $uBVRIz$ photometry, using the  {\textsc ZEBRA} code
\citep{Feldmann06}. This software allows to define 6 basic templates (elliptical, Sbc, Sbd, irregular and two starburst SEDs), 
and to compute log-interpolations between each pair of adjacent templates.
We first applied to the $BVz$ magnitudes the offset derived within the COSMOS survey by \cite{capak07}, that is +0.19 ($B$), +0.04 ($V$), -0.04 ($z$).
We then convolved the stellar spectra from the Pickles' library \citep{Pickles} with the transmission curves used for each filter, and derived the offsets for the other filters, 
that is: 0.0 ($u$), 0.0 ($R$), +0.05 ($I$). Fig.~\ref{fig:stars} shows the comparison 
of the model colors with those derived for the stars in our catalogs, after the above offsets were applied.
We also verified that the offsets derived in this way are consistent with those obtained by running {\textsc ZEBRA}  in the so-called photometry-check mode, 
that allows to compute magnitude offsets minimizing the average residuals of observed versus  template magnitudes.

\begin{figure}
 \centering
 \includegraphics[width=7 cm,angle=270]{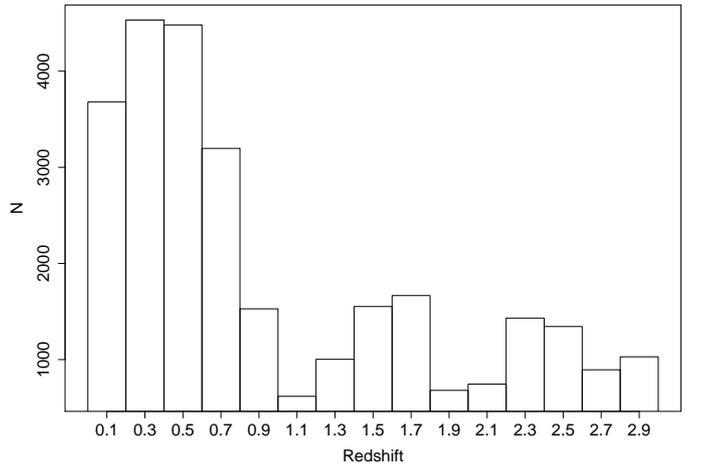}
\caption{Distribution of the photometric redshifts computed from the $uBVRIz$ data for $R < 25$ mag.}
 \label{fig:zhisto}
\end{figure}

We then removed from the catalog those galaxies with a photometric redshift $z_{\rm ph} >3$ and $\sigma_z/(1+z) > 0.1$, where $\sigma_z$  
was derived from the 68\%--level errors computed in {\textsc ZEBRA}. The distribution of the so-obtained photometric
redshifts is displayed in Fig.~\ref{fig:zhisto}.
The accuracy of the so obtained photometric redshifts was derived from the comparison with the SDSS photometric redshifts of galaxies in the SDSS (DR7), 
whose  rms error is $\sim 0.025$ for $r' < 20$ mag. We derived (Fig.~\ref{fig:sdss}) a systematic offset $\Delta z / (1+z) = 0.003$ and an rms error 
$\sigma \Delta z / (1+z) = 0.07$.
As a further check, we extracted from our catalog  those galaxies classified by ZEBRA as early-type, with $R < 23$ mag and $|z_{\rm phot}-0.187| = 0.1$.
As expected, such galaxies define a red sequence (Fig.\ref{fig:redseq}):  this was fitted as $V-R = a + b R$, with $a=0.5$, $b=-6 \times 10^{-3}$.

\section{Shape measurement }
\label{sec:ksbex}

% Short summary of KSB formulae

Ellipticities of galaxies were estimated using the KSB approach \citep{kaiser97}: even if such algorithm does not
allow to achieve accurate measurements of very low shear signals, $\gamma \lesssim 10^{-3}$, it is nevertheless adequate in the case
of weak lensing by clusters, as discussed e.g. by \citet{Gill09} and \citet{Romano10}.

In our KSB implementation, the \textsc{SExtractor} software  was modified to compute all the relevant
quantities, namely the raw ellipticity $e$, the smear
polarizability $P^{\rm sm}$, and the shear polarizability $P^{\rm sh}$. The centers of detected sources
were measured using the windowed centroids in SExtractor.

The KSB approach assumes that the PSF can be described as the sum of an isotropic component  (simulating the effect
of seeing)  and an anisotropic part.
The correction of the observed ellipticity $e_{\rm obs}$ for the anisotropic part is computed as:
\begin{equation}
 e_{\rm aniso} = e_{\rm obs} - P^{\rm sm} p,
\end{equation}
where (starred terms indicate that they are derived from measurement of stars):
\begin{equation}
p = e^*_{\rm obs}/P^{\rm sm*}.
\end{equation}
The intrinsic ellipticity $e$ of a galaxy, and the reduced shear, $g=\gamma/(1-\kappa)$, are then related by:
 \begin{equation}
 e_{\rm aniso} = e +P^{\gamma} g   \label{eq:ellipticity}
\end{equation}
The term $P^{\gamma}$, introduced by \citet{kaiser97} as the \textit{pre--seeing shear polarizability}, describes the effect of  seeing and is defined to be:
\begin{equation}
 P^{\gamma} = P^{\rm sh} - P^{\rm sm} \frac{P^{\rm sh*}}{P^{\rm sm*}} \equiv  P^{\rm sh} - P^{\rm sm} q.
\end{equation}

The final output of the pipeline is the quantity $e_{\rm iso} =
e_{\rm aniso}/P^\gamma$, from which the average reduced shear,
$\left\langle g\right\rangle  = \left\langle e_{\rm
iso}\right\rangle $, provided that the average intrinsic ellipticity
vanishes, $\left\langle e\right\rangle  = 0$.

\begin{figure}
 \centering
 \includegraphics[width=7 cm,angle=270]{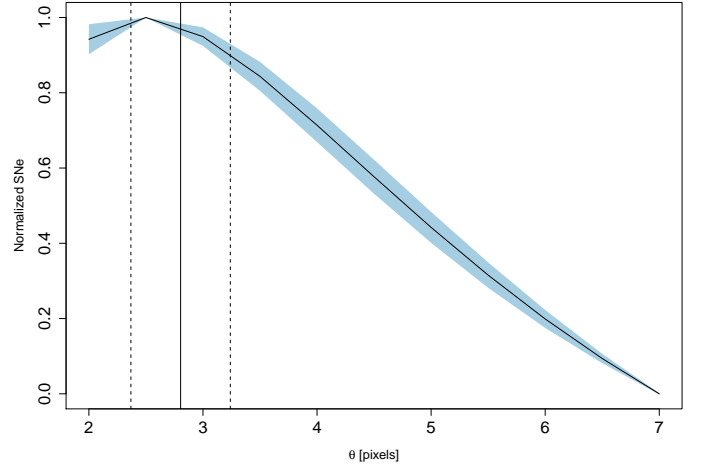}
 % apmass.png: 970x797 pixel, 85dpi, 28.99x23.82 cm, bb=0 0 822 675
 \caption{SNe is displayed as a function of the window function size, $\theta$, used to measure ellipticities.
For display purposes, galaxies
were selected to have the same value of $\theta_{\rm max}$, and SNe was normalized so that min(SNe)=0, max(SNe)=1.
The vertical lines
indicate the average (solid) and standard deviation (dashed) of FLUX\_RADIUS for the same galaxies. }
 \label{fig:sne}
\end{figure}

\begin{figure}
 \centering
 \includegraphics[width=7 cm,angle=270]{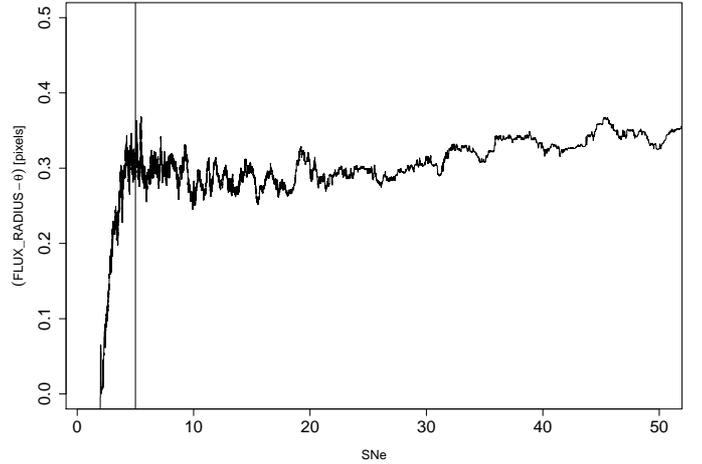}
 % apmass.png: 970x797 pixel, 85dpi, 28.99x23.82 cm, bb=0 0 822 675
 \caption{Running median of FLUX\_RADIUS-$\theta_{\rm max}$ as a function of SNe. The vertical line shows the limit chosen for the
selection of background galaxies.}
 \label{fig:deltar}
\end{figure}

The ellipticity calculation is done by using a window
function in order to suppress the outer, noisy part of a galaxy: the function
is usually chosen to be Gaussian with size $\theta$.
 The size of the window
function is commonly  taken as the radius containing 50\% of the total flux of
the galaxy (as given by e.g. the FLUX\_RADIUS parameter in \textsc{SExtractor}). In our
case, we proceed as follows. We define a set of bins with $\theta$ varying between 2 and 10 pixels (sources with smaller and larger sizes are rejected in our analysis), and a step of 0.5 pixel. For each bin we compute $e_{\rm obs}$, $P^{\rm sh}$ and $P^{\rm sm}$, and
the ellipticity signal to noise ratio defined by Eq. 16 in \citet{erben01}:
\begin{equation}
 {\rm SNe}(\theta) = \frac{\int I(\theta) W(|\theta|)d^2\theta}{\sigma_{\rm sky}\sqrt{\int W^2(|\theta|)d^2\theta}}.
\end{equation}
The optimal size of the window function, $\theta_{\rm max}$, is then defined as the value that maximizes SNe.
Fig.\ref{fig:sne} shows the typical
trend of SNe, which was normalized for display purposes, as a function of $\theta$. It can be seen (Fig. ~\ref{fig:deltar}) that, on average, there is a constant offset between $\theta_{\rm max}$ and FLUX\_RADIUS. Below SNe $\sim$ 5, FLUX\_RADIUS starts to decrease: this provides an estimate of limit on SNe, below which the shape measurement is not meaningful any more.

\begin{figure}
 \centering
 \includegraphics[width=9.5 cm, bb=0 50 612 700,clip]{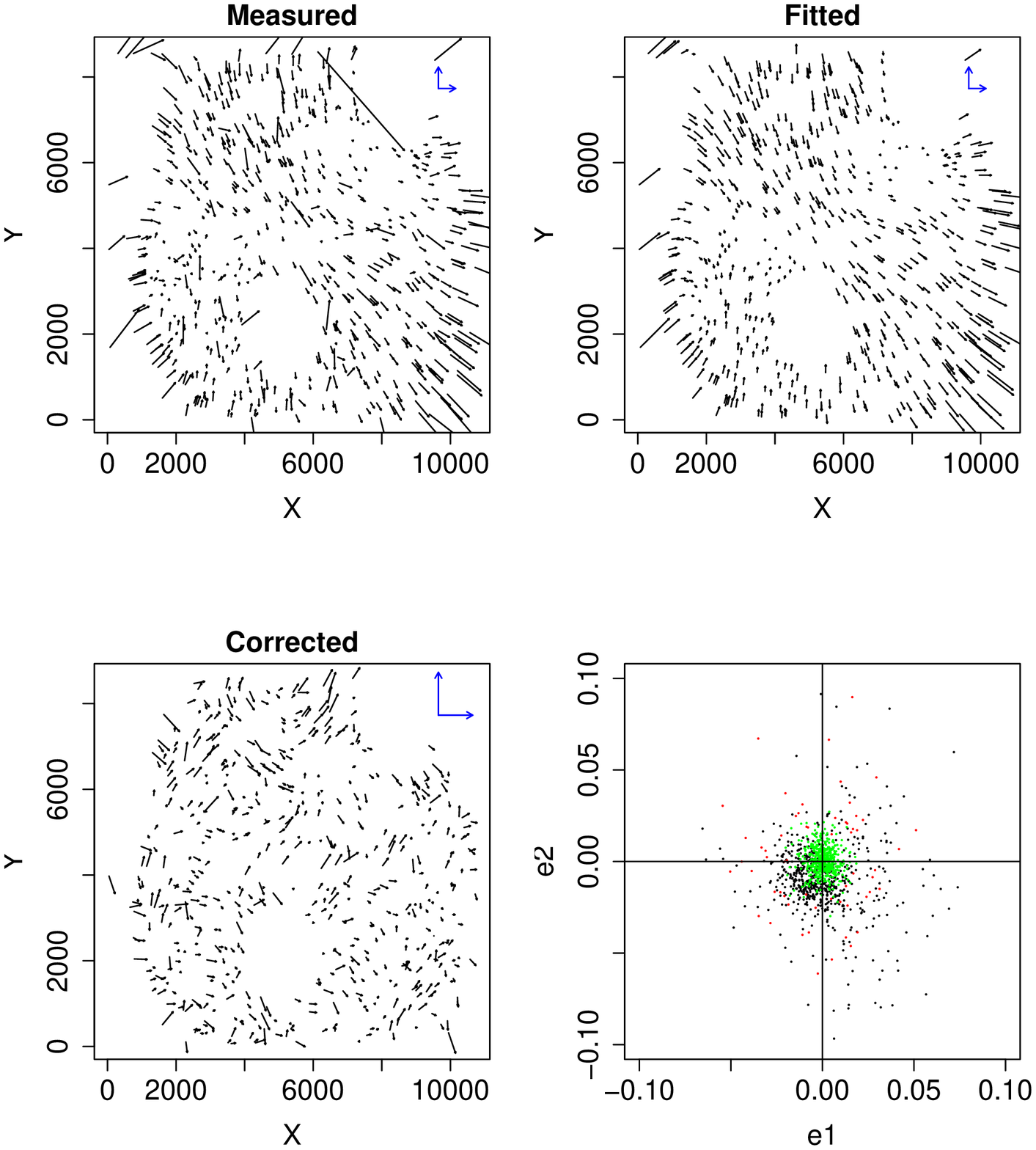}
 \includegraphics[width=9.5cm, bb=0 50 612 360,clip]{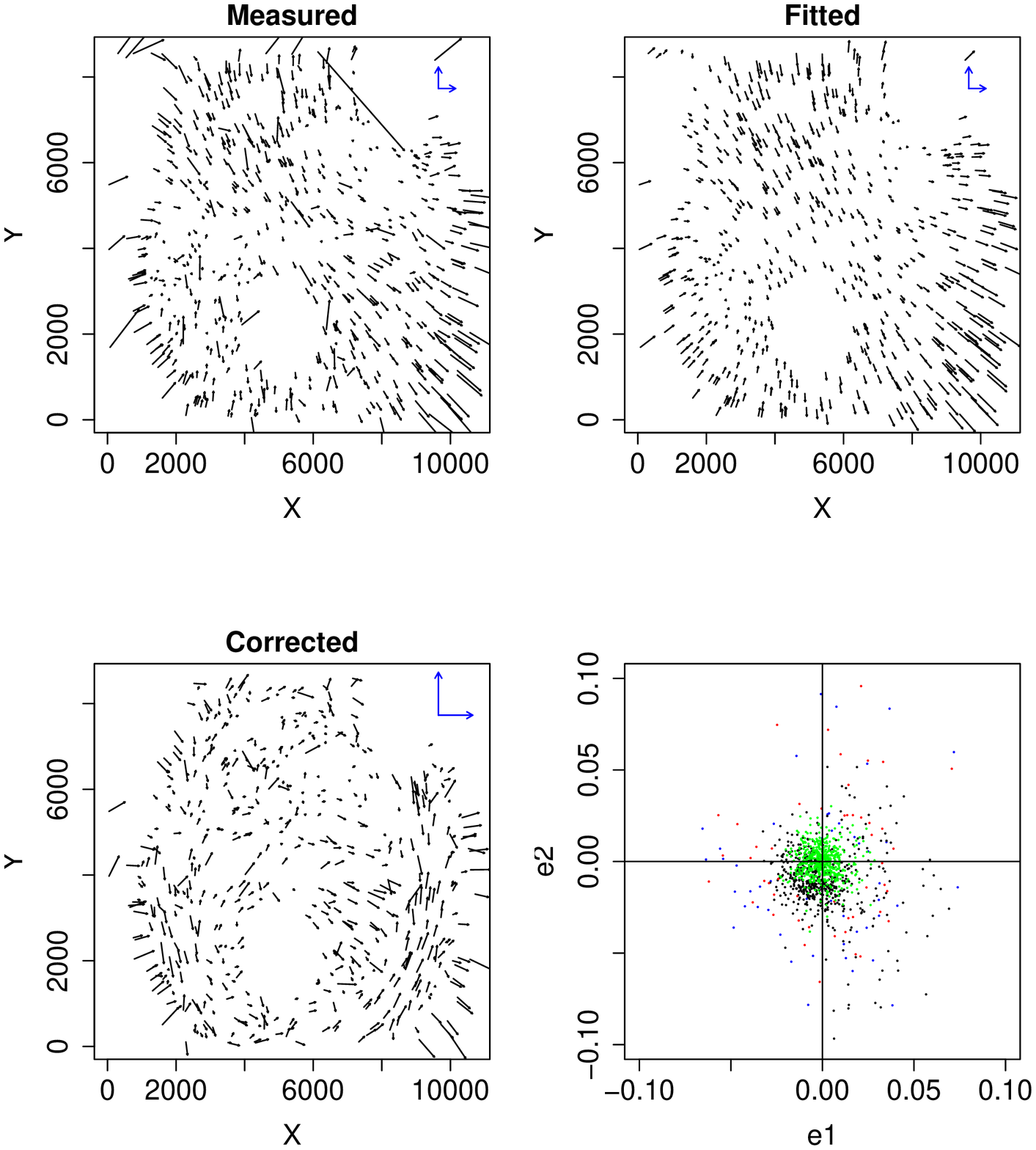}
 % psmap_pl3.ps: 612x792 pixel, 72dpi, 21.59x27.94 cm, bb=0 0 612 360

 \caption{PSF anisotropy correction derived with the GAM algorithm: the first three panels show the ellipticity pattern (measured, fitted and residuals; X and Y are in pixels). The scale is displayed by the arrows in the upper right part of each panel ($e=0.05$).
In the next panel, black dots are the measured values, green dots are after the correction; values rejected during fitting are marked in red. The last row shows for comparison the corrected ellipticities obtained using for the fit a polynomial of order 3.}
 \label{fig:psfmap}
\end{figure}

The terms $p$ and $q$, derived from stars, must be evaluated at each galaxy position: this is usually done fitting them by a polynomial 
\citep[see e.g.][]{Radovich08}, whose order must be chosen to fit the observed trend, without overfitting.  
The usage of the window function introduces a calibration factor, which is compensated by the $P^\gamma$ term.  This implies that stellar terms 
must be computed and fitted with the same value of $\theta$  used for each galaxy \citep{Hoekstra98}.
An alternative approach, not based on a constant (and somehow arbitrary) order polynomial, is given e.g.
by Generalized Additive Models:  we found that the implementation  in R (function \textsc{GAM} in the mgcv library) provides good results.
Fig.~\ref{fig:psfmap} shows fitting and residuals of the anisotropic PSF component: from the comparison between
the results obtained with polynomial and GAM fitting we see that in the latter case we obtain lower residuals, in particular
in the borders of the image.
To quantify the improvement compared to the usage of the polynomial, we obtain 
$\left\langle e_{\rm aniso,1} \right\rangle  = (1 \pm 5) \times 10^{-4}$,  
$\left\langle e_{\rm aniso,2} \right\rangle  = (-2 \pm 9) \times 10^{-4} $ 
with a polynomial of order 3, 
 $ \left\langle e_{\rm aniso,1}\right\rangle  = (2 \pm 4) \times 10^{-4} $,  
$\left\langle e_{\rm aniso,2}\right\rangle  = (1 \pm 7) \times 10^{-4} $ 
with the GAM algorithm.
The values of the fitted terms $p$ and $q$ at the positions of the galaxies are predicted by \textsc{GAM}, that also provides an estimate of the 
standard errors of the predictions, $\Delta p$ and $\Delta q$. From error propagation, the uncertainty on $e_{\rm iso}$ was computed as:
\begin{equation}
\Delta e_{\rm iso}^2 = (\Delta e_{\rm aniso}/P^\gamma)^2 + (e_{\rm aniso} (P^\gamma)^{-2} \Delta P^\gamma)^2,
\end{equation}
where $(\Delta e_{\rm aniso})^2 = (\Delta e_{\rm obs})^2 + (P^{\rm sm} \Delta p)^2$ and $(\Delta P^\gamma)^2 =  (P^{\rm sm} \Delta q)^2$; uncertainties on the measured values of $P^{\rm sm}$ and $P^{\rm sh}$ were not considered.

For each galaxy, a weight is defined as:
\begin{equation}
w=\frac{1}{\Delta {e_0}^2  +\Delta e_{\rm iso}^2 }, \label{eq:well}
\end{equation}
where $\Delta {e_0} \sim 0.3$ is the typical intrinsic rms of galaxy ellipticities.

% de1acorr[xs]=abs(de1[xs]) + abs(Psm11_rg[xs,i]*e1Psm11f$se.fit)
%  de2acorr[xs]=abs(de2[xs]) + abs(Psm22_rg[xs,i]*e2Psm22f$se.fit)
% de1iso<-abs(de1acorr/Pgamma) + abs(e1acorr/(Pgamma*Pgamma)*dPgamma)
%de2iso<-abs(de2acorr/Pgamma) + abs(e2acorr/(Pgamma*Pgamma)*dPgamma)

%w<-1./(de0*de0+(de1iso*de1iso+de2iso*de2iso))

\subsection{Star-galaxy classification}
\label{sec:sgclass}

Stars and galaxies were separated in the magnitude (MAG\_AUTO) vs. size plot. Instead of using e.g.
FLUX\_RADIUS as the estimator of size, we used the quantity $\delta$=MU\_MAX-MAG\_AUTO, where  MU\_MAX is
the peak surface brightness above background. Saturated stars were found in the locus of sources with constant MU\_MAX;
in the $\delta$ vs. MAG\_AUTO plot, stars are identified as sources in the vertical branch.
Sources with $\delta$ lower than stars were classified as spurious detections.
In addition, we rejected those sources for which  $\delta$ is  $\sim 2\sigma$ larger than the median value. This is to
exclude from the sample of stars used to compute the PSF correction terms, those sources for which the shape measurement may
be wrong due to close blended sources, noise, etc.

We further excluded  those galaxies with $w<1$ or SNe $<5$, for which the ellipticity measurement is not meaningful.

\subsection {Error estimate in shear and mass measurement}
\label{sec:simulations}

In the case of  faint galaxies used for the weak lensing analysis, the
ellipticity is underestimated due to noise.
Such effect is not included in the  $P^\gamma$ term,  which can be only computed
on stars with high signal to noise ratio.
\citet{schrabback10} proposed the following parametrization for such bias, as
a function of the signal to noise:
\begin{equation}
m = \frac{e_k-e_m}{e_m} = a * ({\rm SNe})^b,
\end{equation}
where $e_m$ and $e_k$ are the ellipticities before and after the correction respectively.
Such parameters were derived using the STEP1 \citep{step1} and STEP2 \citep{step2} simulations, where
 both PSF and shear are constant for each simulated image.
We obtain $a=-0.1$ and $b=-0.45$, which corresponds to a bias $m$ changing from $\sim 5\%$ for
SNe =5 to $< 2\%$ for SNe $>50$.
After such correction was applied, we computed again the average shear from
the STEP1 and STEP2 simulations, and obtained a typical bias of $\sim$ 3\% for SNe $=$ 5.

We then estimated the accuracy on the mass that can be obtained from an image with the same noise and depth as
in the $R-$band SUBARU image.
To this end we dropped the assumption on constant shear, and produced more realistic
simulations: the effect on galaxy shapes by weak lensing from a galaxy cluster was produced
using the \textsc{SHUFF} code, that will be described in a separate paper (Huang et al., in preparation).
To summarize, the code takes as input a catalog of galaxies produced
by the \textsc{Stuff} tool; it computes the shear produced by a  standard mass
profile (e.g. Navarro-Frenk-White, NFW hereafter) and applies it to the ellipticities of the galaxies behind
the cluster. Such catalog is then used in the \textsc{SkyMaker} software, configured with the telescope parameters
suitable for the SUBARU telescope and with the  exposure time of the $R-$band image, producing a simulated image;
the background rms of such image was set to be as close as possible to that of the real image.
We considered for the lens a range of masses at
$\log M_{\rm vir}/M_\odot = 13.5, 14, 14.5, 15.0$, and a NFW mass profile with
 $c_{\rm vir} = 6$. Each simulation was repeated 50 times for each mass value, randomly changing the morphology,
position and redshift of the galaxies.

For each of these images, we run our lensing pipeline with the same configuration
used for the real data. The  density of background galaxies used for the lensing analysis was $\sim 20$ gals arcmin$^{-2}$.
The fit of the mass was done as described in \citet{Radovich08} and
\citet{Romano10}: the expressions for  the radial dependence of tangential shear $\gamma_{T}$ derived by
\citet{Bartelmann96} and \citet{Wright00} were used, and the NFW parameters ($M_{\rm vir}$, $c_{\rm vir}$) were derived
using a maximum  likelihood approach.  In addition, the 2D projected mass can be derived in a non-parametric way by
aperture densitometry, where the mass profile of the cluster is computed by the $\zeta$ statistics \citep{fahlman, apj...497l..61c}:
\begin{eqnarray}
 \zeta (\theta_1) =
  \bar{\kappa} (\theta \le \theta_1) - \bar{\kappa} (\theta_2 < \theta \le \theta_{\rm out}) =
2  \int^{\theta_2}_{\theta_1} \left\langle \gamma_T \right\rangle d \ln \theta \\ \nonumber
+  \frac{2}{1 - (\theta_2 /\theta_{\rm out})^2} \int^{\theta_{\rm out}}_{\theta_2} \left\langle \gamma_T \right\rangle d \ln \theta. \label{eq:apmass}
\end{eqnarray}
The mass is estimated as $M_{\rm ap}(\theta_1) = \pi \theta_1^2 \zeta (\theta_1) \Sigma_{\rm crit}$, and $\theta_{\rm out}$ is chosen so that  $\bar{\kappa} (\theta_2, \theta_{\rm out}) \sim 0$.

The average errors on mass estimate obtained in such way are displayed in  Table ~\ref{tab:msimul}, showing that masses can be estimated within an uncertainty 
of $<$ 20\% for $M \ge 10^{14} M_\odot$.
Such accuracy only includes the contribute due to shape measurement and mass fitting method, but it
does not include the  uncertainty due to the selection of the lensed galaxies.

Finally, the masses derived by aperture densitometry are $\sim$ 1.3 higher than those obtained by mass fitting: this is in agreement with \citet{Okabe10},  
who find $M_{\rm 2D}/M_{\rm 3D}=1.34$ for virial overdensity.

\begin{table}
 \begin{center}
% use packages: array
\caption{Masses  derived by simulations with NFW model fitting ($M_{\rm 3D}$) and aperture
densitometry ($M_{\rm 2D}$). Input masses used for each simulation are given in the first column 
($ M_{\rm in} $).\label{tab:msimul}}

\begin{tabular}[c]{lll}
\hline\hline

$ M_{\rm in} $ &$M_{\rm 3D}$  & $M_{\rm 2D}/M_{\rm 3D}$ \\
$10^{14} M_\odot$ & $10^{14} M_\odot$\\
\hline
0.316 & $0.31 \pm 0.13$  & $ 1.3 \pm 0.9$\\
1.00  & $1.01 \pm 0.24$  & $ 1.4 \pm 0.4$ \\
3.16  & $ 3.12 \pm 0.36$ &  $ 1.4 \pm 0.3$ \\
10.0  & $ 10.1 \pm 0.5 $ &  $ 1.3 \pm 0.1$ \\
\hline
 \end{tabular}
 \end{center}

\end{table}

\begin{figure*}
 \centering
\includegraphics[width=6 cm,angle=270]{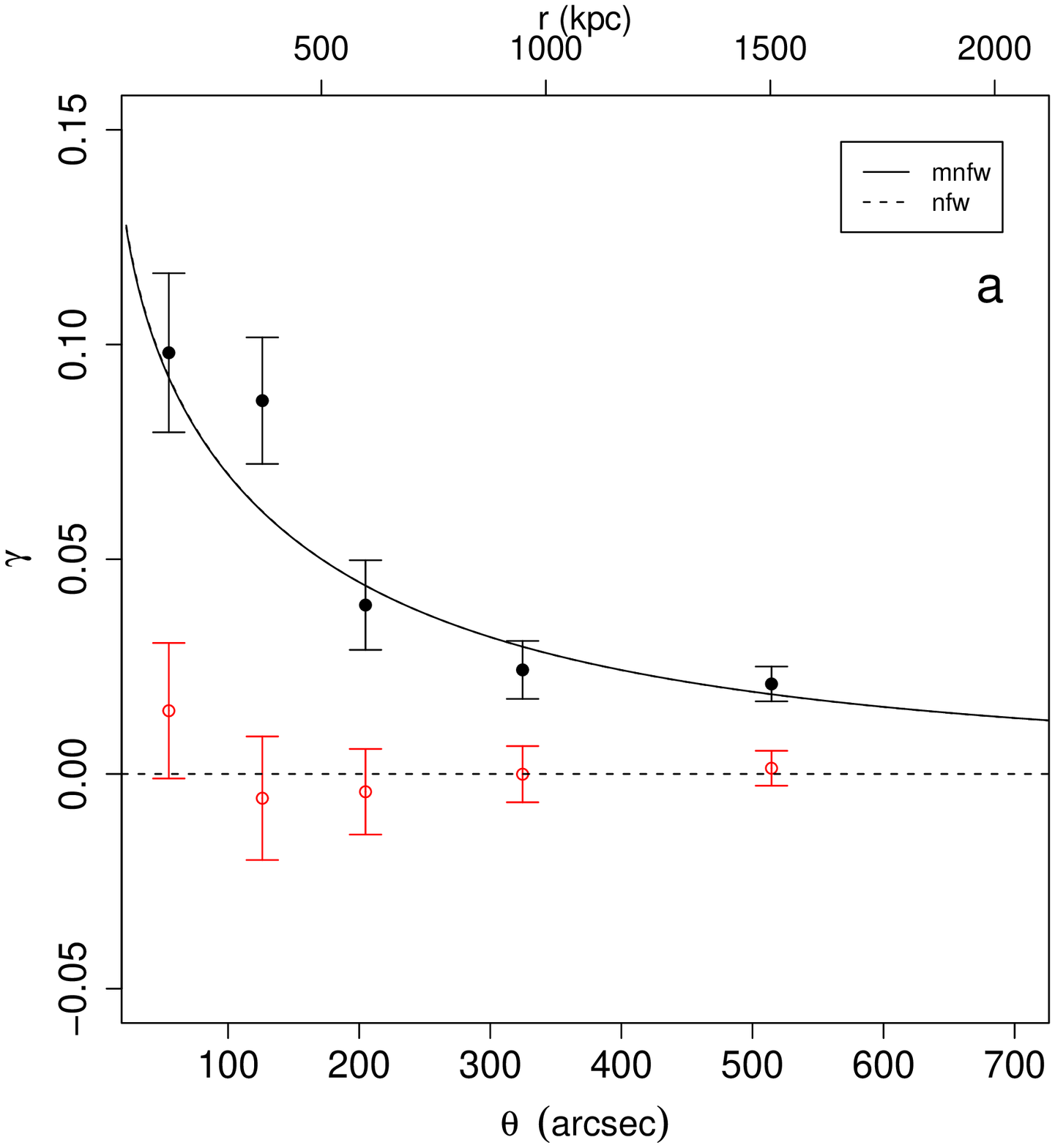}
\includegraphics[width=6 cm,angle=270]{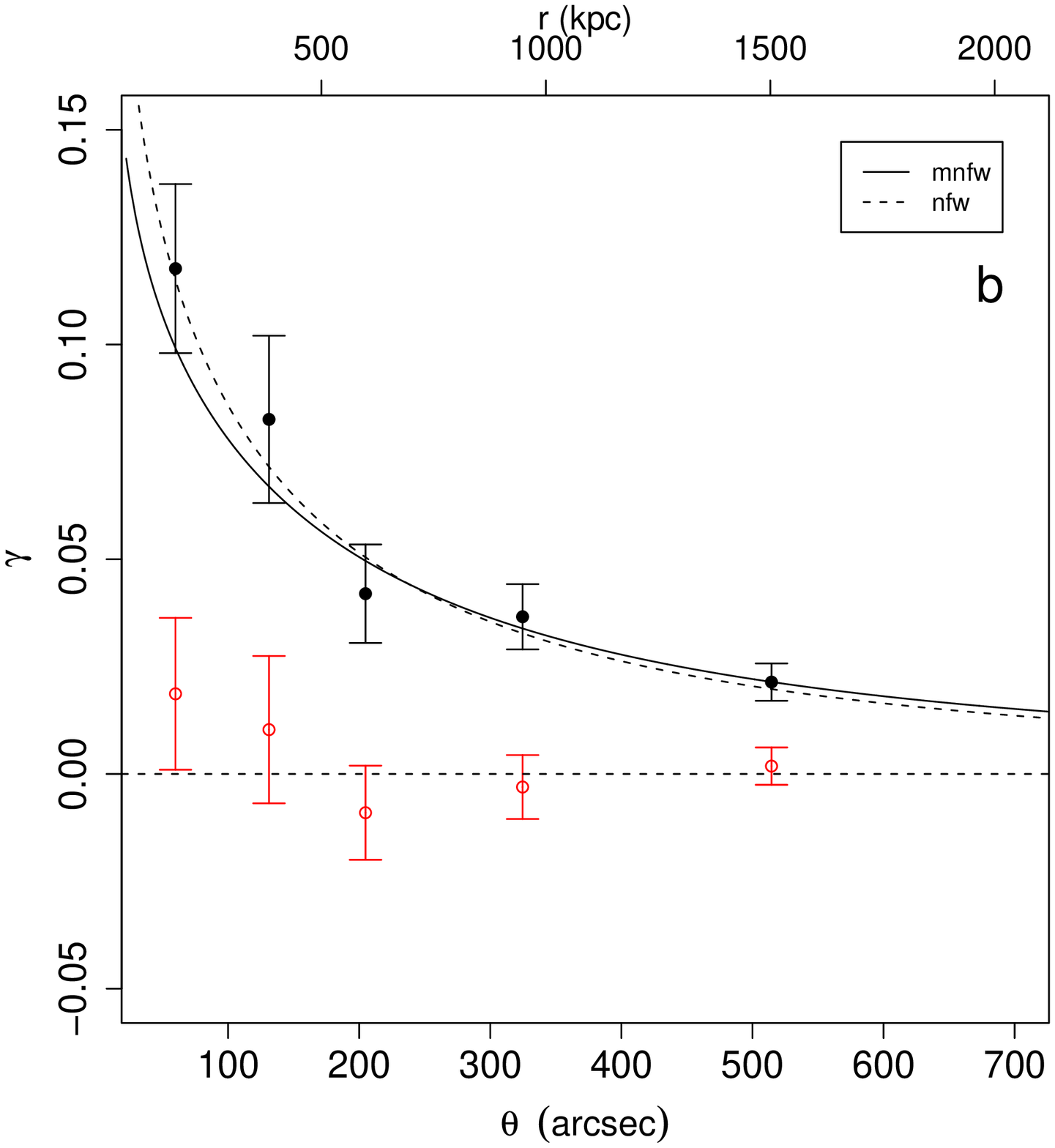}
\includegraphics[width=6 cm,angle=270]{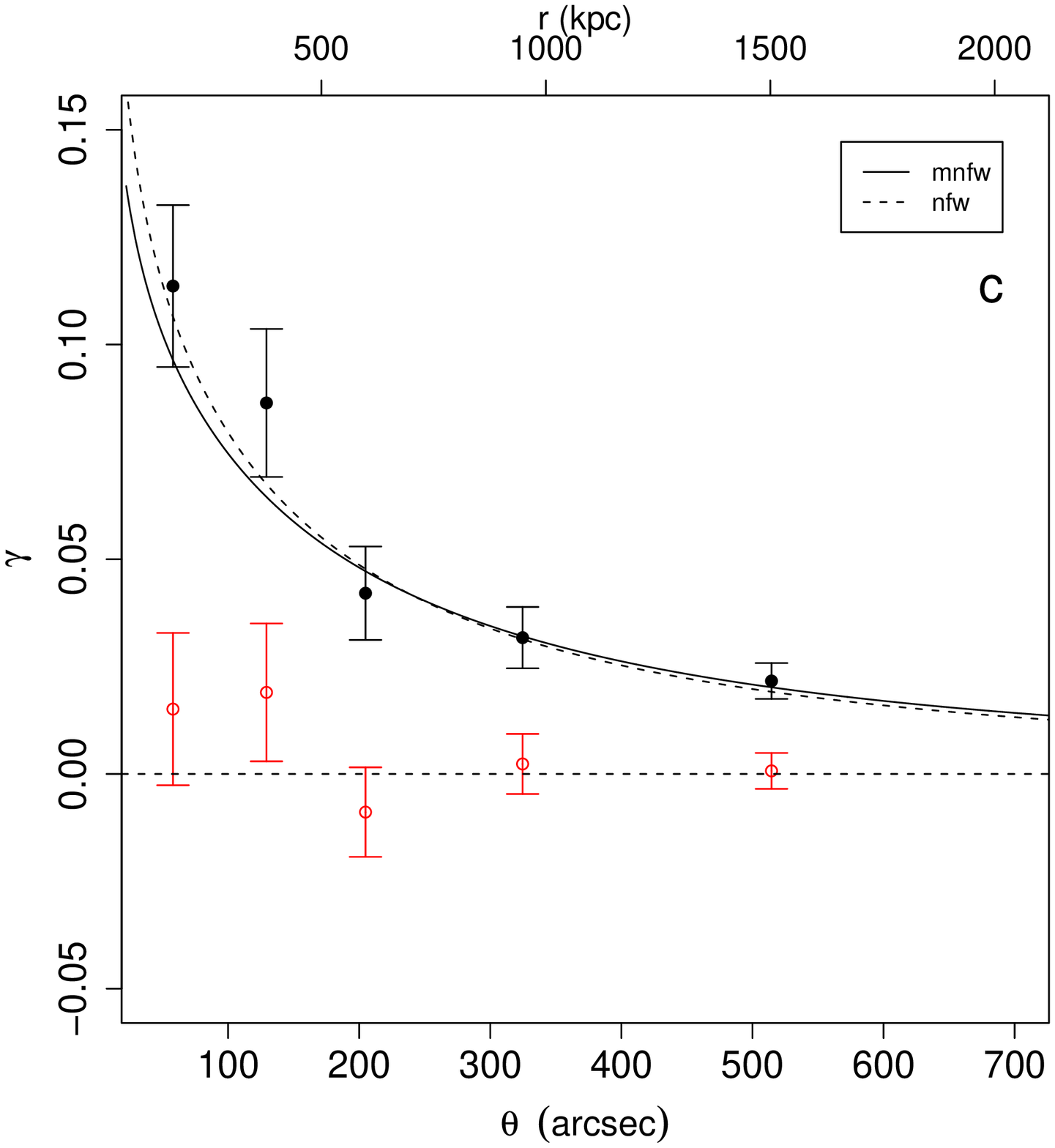}
\includegraphics[width=6 cm,angle=270]{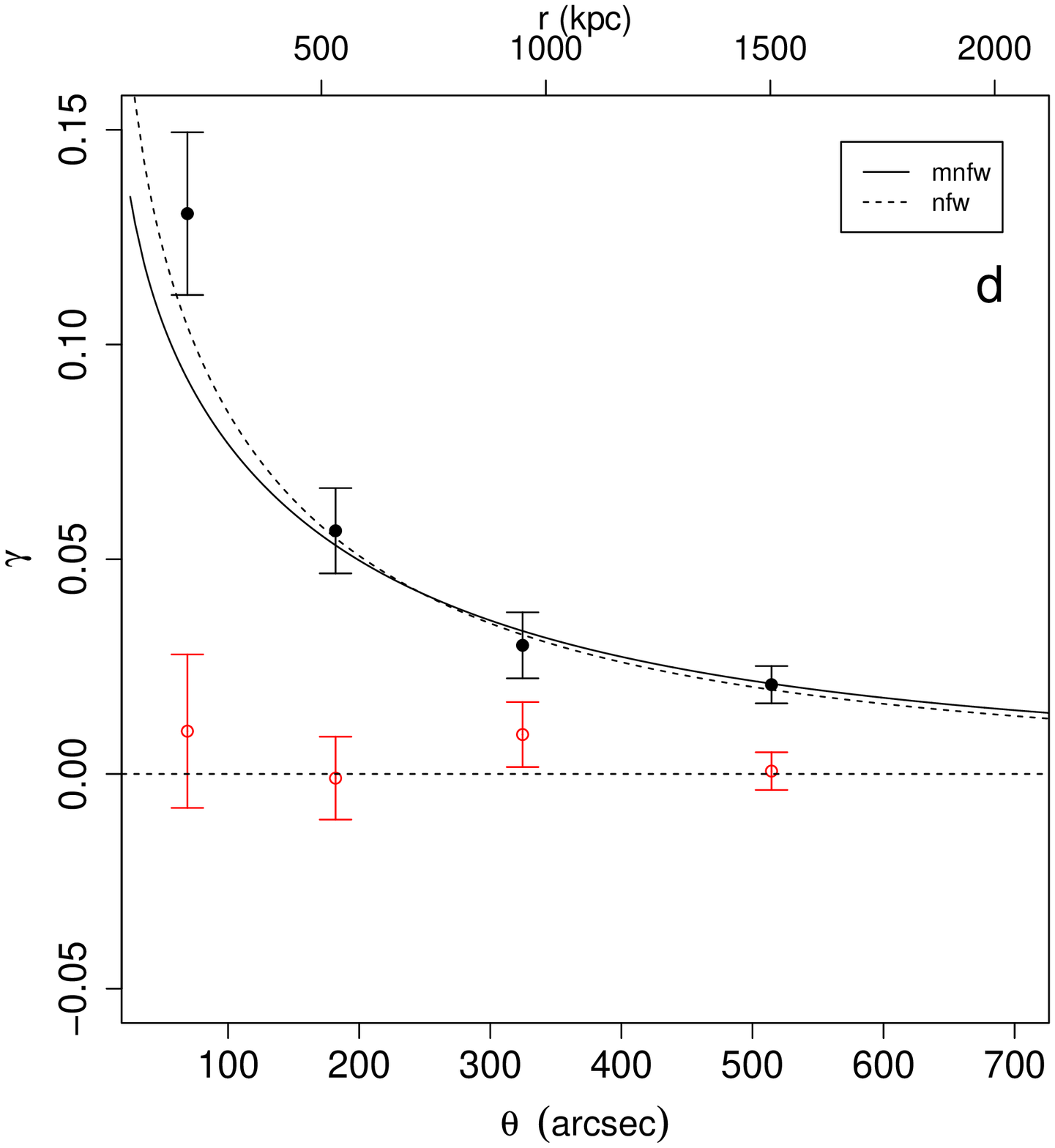}

 \caption{Shear profiles obtained with the different selection methods (see Table~\ref{tab:nfw}, where the parameters derived for each model are given).
The fitting was done using the maximum likelihood approach. Binned points are shown for display purposes only. In panel {\em a}, the curves obtained by the two models (nfw/mnfw) overlap.}
 \label{fig:shearprof}
\end{figure*}

\section {Results}
\label{sec:results}

One of the most critical source of systematic errors, which can lead to an
underestimation of the true WL signal, is dilution of the distortion due to the
contamination of the background galaxy catalog by unlensed foreground and
cluster member galaxies \citep[see e.g.][]{Broadh05}.
The dilution effect increases as the cluster-centric distance decreases because
the number density of cluster galaxies that contaminate the faint galaxy
catalog is expected to roughly follow the underlying density profile of the
cluster. Thus, correcting for the dilution effect is important to obtain
unbiased, accurate constraints on the cluster parameters and mass profile.

As discussed by \citet{Broadh05,Okabe10,Oguri10}, the selection of background galaxies to be used for the weak lensing analysis can be done taking those galaxies redder than the cluster red sequence. However, such selection
produces a low number density (10 galaxies/arcmin$^2$ in our case), and correspondingly high uncertainties in the derived parameters. 
 In the following, we compare the results obtained by different methods.
We first assumed that no information on the redshift is available, and that photometry from only one band is available
(magnitude cut), or from more than two bands (color selection). Finally, we included in our analysis the  photometric redshifts.
The density of background galaxies is $\sim$ 25-30  galaxies/arcmin$^2$, see Table~\ref{tab:nfw}.

In order to derive the mass, we need to know the critical surface density:
\begin{equation}
\Sigma_{\rm crit} = \frac{c^2}{4\pi G} \frac{D_s}{D_l D_{ls}} = \frac{c^2}{4\pi G} \frac{1}{D_l \beta},
\end{equation}
 $D_{ls}$, $D_s$, and $D_l$ being the angular distances between lens and source, observer and source, and observer
and lens respectively. This quantity should be computed for each lensed galaxy. As the reliability of photometric redshifts
for the faint background galaxies  is not well known, we prefer to adopt the single sheet approximation, where all background galaxies are assumed to lie at the same redshift,  defined as 
$\beta(z_s) = \left\langle \beta(z) \right\rangle $. In the case of the selection based only on magnitude or colors, such value was derived from the COSMOS catalog of photometric redshifts  \citep{capak07}, to which the same cuts used for the Abell 383 catalog are applied. Later on, we computed $\beta(z_s)$ from the photometric redshifts themselves, and compared such two values.

The mass was  computed both by fitting a NFW profile ($M_{\rm 3D} = M_{\rm vir}$), and by aperture densitometry ($M_{\rm 2D}$). In the case of the NFW fit, in addition to a 2-parameters fit (virial mass $M_{\rm vir}$
and concentration $c_{\rm vir}$), we also show (MNFW) the results obtained using the relation proposed
by \citet{Bullock01} between $M_{\rm vir}$, $c_{\rm vir}$ and the cluster redshift $z_{\rm cl}$:
\begin{equation}
c_{\rm vir} =\frac{K}{1+z_{\rm cl}} \left(\frac{M_{\rm vir}}{M_\star}   \right)^\alpha,
\label{eq:cvirmass} 
\end{equation}
with $M_\star= 1.5 \times 10^{13}/h$ $M_{\odot}$, $K = 9$, $\alpha=-0.13$.

The shear profiles obtained from the different methods discussed below are displayed in Fig.~\ref{fig:shearprof}; also displayed are
the average values of tangential and radial components of shear, computed in  bins selected to contain at least 200
galaxies, and centered on the BCG: this is also where the peak of the
X--ray emission is located \citep{Rizza98}. In order to check the possible error introduced by a wrong center, we considered a grid around the position of the
BCG, with a step of 2 arcsec: for each position in the grid, we took it as the center, performed the fit and derived the mass. Within 30 arcsec, we obtain 
that the rms is $\sigma(M_{\rm vir}) < 5\%$. 
The NFW parameters obtained by model fitting, and the reduced  $\chi^2$  computed from the binned average tangential 
shear,  are given in Table~\ref{tab:nfw}.

\begin{figure}
 \centering
 \includegraphics[width=7 cm,angle=270]{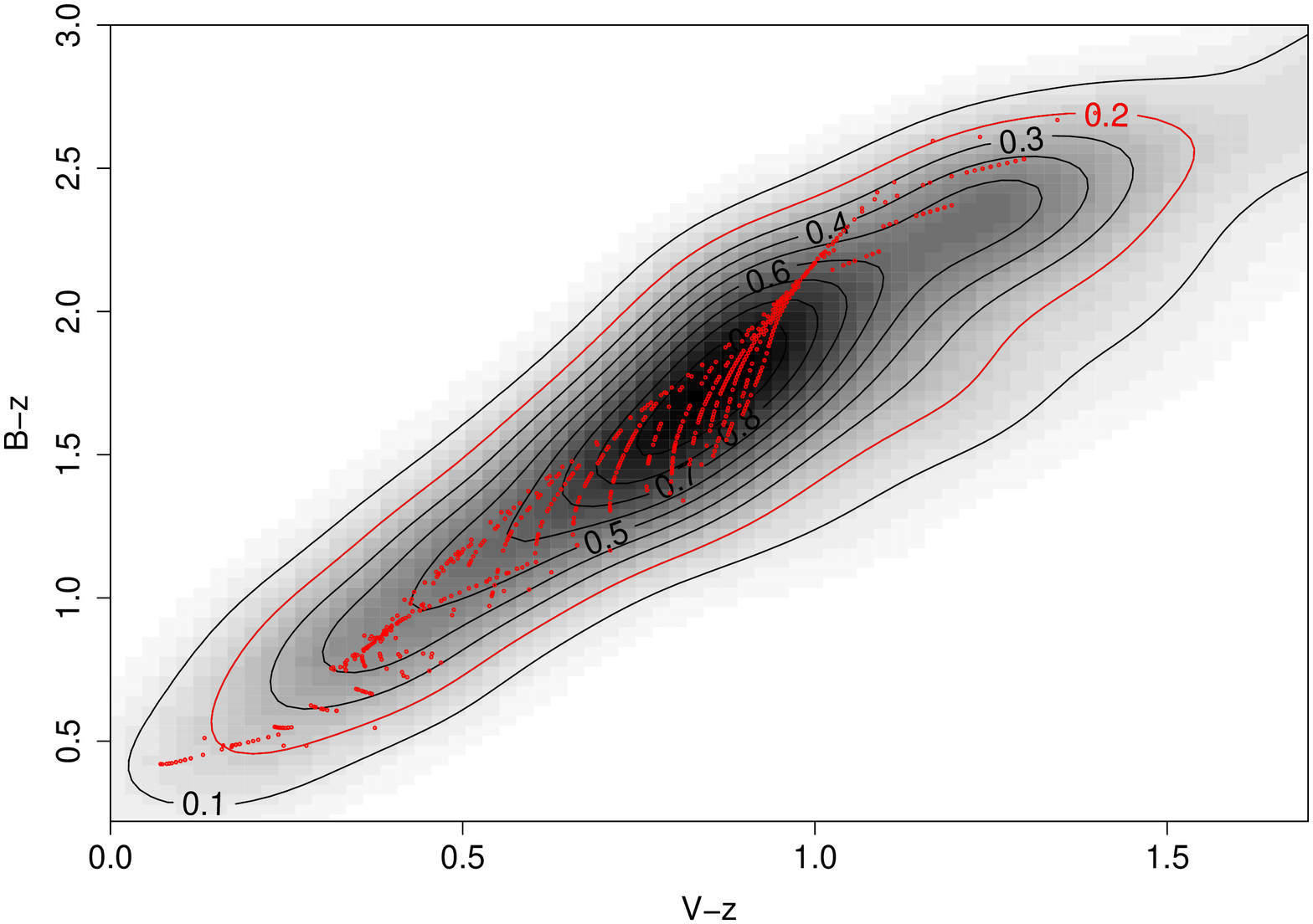}
 \includegraphics[width=7 cm,angle=270]{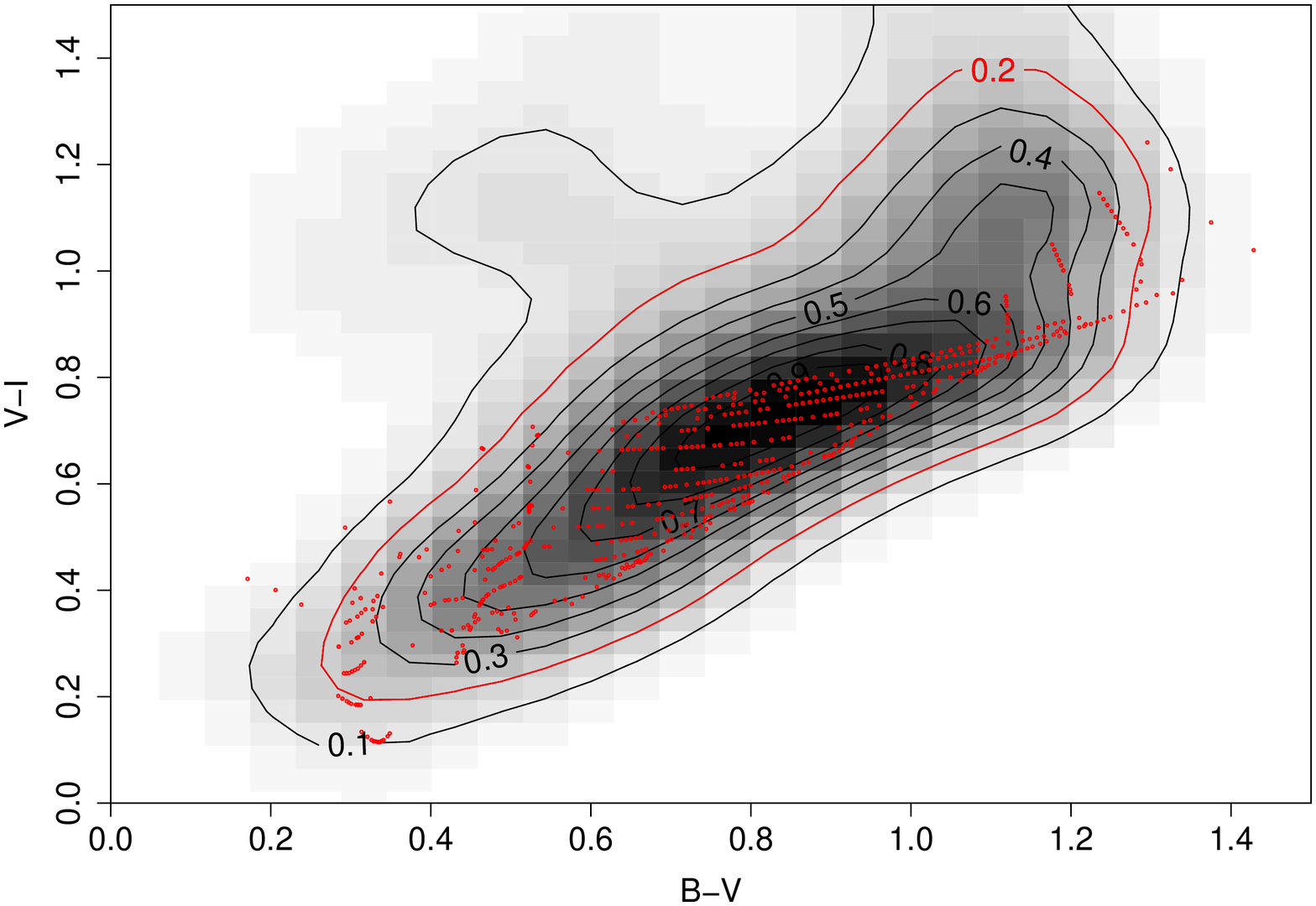}
 % apmass.png: 970x797 pixel, 85dpi, 28.99x23.82 cm, bb=0 0 822 675
 \caption{Color selection of foreground galaxies ($z_{\rm phot}<0.2$). The contour in red is the
density level chosen for the selection; the points display the model colors computed by
ZEBRA for this redshift range. }
 \label{fig:colsel}
\end{figure}

The magnitude cut is the simplest approach as it only requires photometry from the same band in which the lensing measurement is done.  
Taking galaxies in the range 23 $<$ $R$ $<$ 26 mag ({\em a}), produces a sample dominated by faint background galaxies, but the inner regions of the cluster may still present an unknown contamination by cluster galaxies.

To improve the selection, we proceeded as follows. The locus of foreground galaxies was first found, allowing a
better separation of different galaxy populations \citep[see][and references therein]{Medez10}, compared e.g. to methods based
on the selection of only red galaxies.
Here we considered two colors  selections, namely  $B-z$ vs. $V-z$ ({\em b}) and $B-V$ vs. $V-I$ ({\em c}), with  $21 < R < 26$ mag.
For intermediate redshifts ($z \sim 0.2$), foreground and background objects are well separated in such two colors diagrams.
We  explored the possibility to find the best selection criteria based only on the observed colors, without any information
on the redshift of the galaxies.
To this end, we developed a semi-automatic procedure, implemented in the R language \citep{R}. We first selected bright ($R < $ 21 mag) galaxies, which are
expected to be mainly at  $z_{\rm ph} < 0.2$. A kernel density estimate, obtained by the kde2d package in R \citep{venables02} was then applied to these points,
giving the plots displayed in Fig.\ref{fig:colsel}: the normalization was done so that the  value of the maximum density in the binned data was equal to one. The boundary of the foreground galaxy region was then defined by the points within the same density level $l$ (e.g., $l=0.2$).  Such region  was converted to
a polygon using the splancs\footnote{http://www.maths.lancs.ac.uk/\~rowlings/Splancs} package in R, that also allows to select for a given catalog those sources whose colors lie inside or outside the polygon.  A comparison with the model colors obtained in ZEBRA from the convolution of the spectral templates with filter transmission curves, shows that colors inside the area selected in such way are consistent with those expected for galaxies at redshift $<$ 0.2. Galaxies classified as foreground in such way were
therefore excluded from the weak lensing analysis.

We finally used photometric redshifts ({\em d}), both for the selection of background galaxies, defined as those with
$21 < R < 26$ mag, $0.3 < z_{\rm ph} < 3$, and to compute the average value of $\beta$: we obtain in such way
$\beta(z_s) = 0.74$, in good agreement with the value obtained from the COSMOS catalog with the same magnitude and redshift selection, $\beta(z_s)=0.73$. 

The effect of the different selections on the residual presence of cluster galaxies is displayed in Fig.~\ref{fig:rdens}, showing the density of background
galaxies computed in different annuli around the cluster: a clear increase of the density in the inner regions is visible in case {\em a}, which indicates that
magnitude selection alone does not allow to completely remove the contamination by cluster galaxies. Such contamination is greatly reduced by color selection, and
the optimal result is given by photometric redshifts, as expected. As a further check, we also found for each method that the tangential shear signals of the rejected 'foreground/cluster' galaxies average out. 
In the following discussion, we take as reference the results from case {\em d}, which is very close to {\em c} in terms of
uncertainties on fitted parameters, density of background galaxies and  residuals in the radial component of the shear.

It was pointed out in \citet{Hoekstra03} and \citet{Hoekstra10} that large-scale structures along the line of sight provide a source of uncertainty 
on cluster masses derived by weak lensing, which is usually ignored and  increases as a larger radius ($\theta_{\rm max}$) is used in the fitting. 
The uncertainty introduced by such component on the mass estimate can be
$\sim$  10-20\% for a cluster with $M=10^{15} M_\odot$ at $z \sim 0.2$, $\theta_{\rm max} = 10$ arcmin as in our case (see Fig.~6 and 7 in  \citet{Hoekstra03}), 
which is comparable to the uncertainties derived in the fitting.

\begin{table*}
\begin{center}
\caption {Best-fit NFW parameters: for each selection method, in the upper row both $M_{\rm vir}$ and $c_{\rm vir}$ were
taken as free parameters, in the lower row the ($M_{\rm vir}$, $c_{\rm vir}$, $z_{\rm cl}$) relation \citep{Bullock01} was used. 
The reduced $\chi^2$ was computed from the binned values displayed in Fig.~\ref{fig:shearprof}.
Also given are the values of $M_{\rm 200}$:
the larger uncertainties are due to the fact they were derived from the fitted values ($M_{\rm vir}, c_{\rm vir}$). The values of $M_{\rm 2D}$ before and after applying the factor 1.34 (Sect.\ref{sec:simulations}) are displayed in the upper and lower rows, respectively. \label{tab:nfw}}

\begin{tabular}{llllllll}

\hline\hline

Method & $M_{\rm vir}$ & $c_{\rm vir}$ & $r_{\rm vir}$ & $M_{\rm 200}$ & $\chi^2$/d.o.f. & $M_{\rm 2D}$ & Density \\
 & $10^{14} M_\odot$ & & arcsec & $10^{14} M_\odot$ && $10^{14} M_\odot$ & gal/arcmin$^2$ \\
\hline
\\
% mag selection
a & $7.78_{-2.21}^{3.34}$ & $4.73_{-1.49}^{1.99}$ & $698_{-73}^{88}$ &
   $6.44^{7.94}_{-3.23}$ & 1.13  &12.3 $\pm$ 2.98  & 30 \\

\\
  & $7.80_{-1.52}^{1.65}$ & $4.71_{-0.12}^{0.13}$ & $699_{-49}^{46}$ &
    $6.46_{-1.68}^{1.98}$ & 1.13 & \textit{9.20 $\pm$ 2.22} \\
\\
\hline
% color selection DEEP [Z:  V-z B-z]
\\
b &$7.61_{-2.04}^{2.94}$ & $5.77_{-1.69}^{2.33}$ & $693_{-69}^{80}$ &
   $6.42^{6.85}_{-3.21}$ & 0.40  & 8.14 $\pm$ 3.15& 25\\
\\
  & $9.00_{-1.69}^{1.85}$ & $4.62_{-0.11}^{0.13}$ & $733_{-49}^{47}$&
   $7.44_{-1.86}^{2.20}$ &  0.70 & \textit{6.07 $\pm$ 2.35}\\
\\
\hline
% color selection DEEP [Z: B-V V-I]
\\
c & $7.51_{-2.00}^{2.90}$ & $5.40_{-1.52}^{2.13}$ & $690_{-68}^{79}$  &
   $6.30^{6.15}_{-3.08}$ &  0.87 & 10.5 $\pm$ 3.03 & 28\\
\\
  & $8.38_{-1.57}^{1.69}$ & $4.67_{-0.11}^{0.13}$ & $716_{-48}^{45}$  &
  $6.94_{-1.73}^{2.01}$ & 0.61 & \textit{7.82 $\pm$ 2.26}\\
\\
\hline
% phz selection [int. avbeta]
\\
d & $7.54_{-1.91}^{2.66}$ & $5.68_{-1.60}^{2.11}$ & $691.05_{-64}^{73}$ &
 $6.35^{6.45}_{-3.02}$ & 1.1  & 10.2 $\pm$ 2.94  & 25 \\
\\
  & $8.73_{-1.60}^{1.75}$ & $4.64_{-0.11}^{0.12}$ & $725.77_{-47}^{46}$ &
 $7.22_{-1.77}^{2.09}$ & 1.90 & \textit{7.59 $\pm$ 2.19 }\\

\hline

\end{tabular}
\end{center}

\end{table*}

%================

\begin{figure}
 \centering
\includegraphics[width=8 cm]{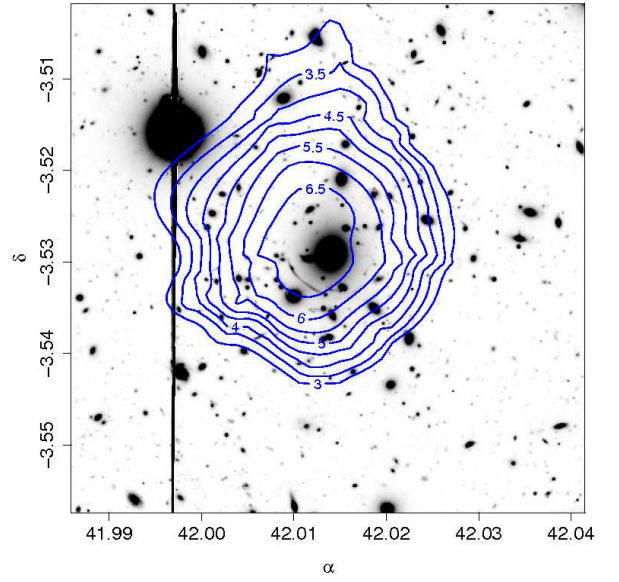}
 \caption{Weak lensing \textit{S-map} showing the mass distribution derived by weak lensing; overlaid is the central region of the Abell 383 field.}
 \label{fig:massdist}
\end{figure}

Also displayed in Table~\ref{tab:nfw} are the projected masses computed by aperture densitometry  from Eq.~\ref{eq:apmass}, at a distance from the cluster center  $r=r_{\rm vir}$, and $\theta_2=900\arcsec$, $\theta_{\rm out}=1000\arcsec$. A good agreement is found between these masses and the values computed from parametric fits, if we take into account the expected ratio  $M_{\rm 2D}/M_{\rm 3D} = 1.34$ (see Sect.~\ref{sec:simulations}).

From the catalog based on photometric redshifts selection (case {\em d}), we finally derived the
 \textit{S-map} introduced by \citet{aa...420...75s}, that is:
 $S=M_{\rm ap}/\sigma_{M_{\rm ap}}$, where
\begin{eqnarray}
M_{\rm ap} & = & \frac{\sum_i  e_{t,i} w_i Q(\vert \theta_i-\theta_0 \vert )}{\sum_i{w_i}}\\
\sigma^2_{M_{\rm ap}} &= &\frac{\sum_i e^2_{i} w^2_i Q^2(\vert \theta_i-\theta_0 \vert)}{2(\sum_i{w_i})^2}, \nonumber
\end{eqnarray}
The map was obtained by defining a grid of points along the image;
the tangential components $e_{t,i}$ of the lensed galaxy ellipticities were computed taking as center each point in such grid.
The weight $w_i$  was defined in Eq.~\ref{eq:well}, and $Q$ is a Gaussian function as in \citet{Radovich08}:
 \begin{equation}
Q(\vert \theta-\theta_0 \vert) = \frac{1}{\pi \theta^2_s} \exp \left( -\frac{(\theta-\theta_0)^2}{\theta^2_s}\right),
\end{equation}
where $\theta_0$ and $\theta_s$ are the center and size of the aperture ($\theta_s \sim 1.5$ arcmin).
The S-map is displayed in Fig.~\ref{fig:massdist}, showing a quite circular mass distribution centered on the BCG.

\begin{figure}
 \centering
 \includegraphics[width=7 cm,angle=270]{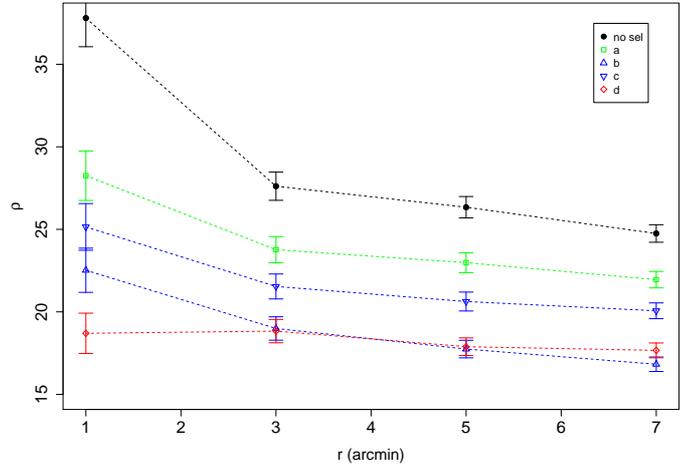}
 % apmass.png: 970x797 pixel, 85dpi, 28.99x23.82 cm, bb=0 0 822 675
 \caption{Density (gal/arcmin$^2$) of background galaxies used for the lensing analysis, as a function of the distance from the cluster, for the different selection methods here considered. The case with no selection is also displayed for comparison.}
 \label{fig:rdens}
\end{figure}

\begin{table}
\begin{center}
\caption{Luminosity function parameters, and uncertainties.
$\Phi _{\star }$ is in units of deg$^{-2}$.
\label{tab:LF}}

\begin{tabular}{llll}
\hline\hline
& Best-fit & Standard error \\
\hline
$\alpha$ & -1.06 & 0.07 &   \\
$m_{\star}$    & 18.32 & 0.39 &   \\
$\Phi_{\star}/10^3$ & 1.273 & 0.374 &  \\
\hline
\end{tabular}

\end{center}
\end{table}

\section{Discussion}
\label{sec:discussion}

Several mass measurements of this cluster are available in literature, based on different data and/or methods.
\citet{Schmidt07} used  Chandra data and modeled the dark matter halo by a generalized NFW profile, obtaining a mass value
$M_{\rm vir}=9.16^{+1.89}_{-1.85} \times 10^{14} M_\odot$ and a concentration value $c_{\rm vir}=5.08^{+0.55}_{-1.03}$.\\

A weak lensing analysis of Abell 383 was done by \citet{Bardeau07} using CFH12K data in the $B$, $R$, $I$ filters. For the shape measurements
they used a Bayesian method implemented into the IM2SHAPE software. To retrieve the weak lensing signal they selected the
 background galaxies as those within $21.6<R<24.9$ mag and  ($R-I$)$\gtrsim 0.7$, obtaining a number density
of $\sim 10$ gal arcmin$^{-2}$. Their fit of  the shear profile by a NFW profile gave as result a mass of $M_{200}=4.19\pm1.46\times h^{-1}_{70} 10^{14} M_\odot$ at  $r_{200}=1.32\pm0.17 h^{-1}_{70}$ Mpc and a concentration value of $c=2.62\pm0.69$. \\

Another weak lensing mass estimate of Abell 383 from CHF12K data was obtained by \citet{hoekstra07}  using
two bands, $B$ (7200 s) and $R$ (4800 s). The background sample was selected by a magnitude cut $21<R<24.5$, from
which cluster red sequence galaxies were discarded. The remaining contamination was estimated from the stacking of
several clusters by assuming that fraction of cluster galaxies $f_{gc}$ was a function of radius $\propto r^{-1}$.
This function was used to correct the tangential shear measurements. As discussed in \citet{Okabe10}, this kind
of calibration method does not allow to perform an unbiased cluster-by-cluster correction.
Assuming a NFW profile, the fitted virial mass of Abell 383 was $M_{\rm vir} = 2.8^{1.6}_{-1.5} \times h^{-1} 10^{14} M_\odot$.

Abell 383 belongs to the clusters sample selected for the Local Cluster Substructure Survey (LoCuSS) project (P.I. G. Smith).
Within this project, a weak lensing analysis of this cluster has been recently performed by \citet{Okabe10} using SUBARU data in two filters, $i'$
(36 minutes) and $V$ (30 minutes). In addition to a magnitude cut $22<i<26$ mag, they used the color information to select galaxies redder and
bluer than the cluster red sequence. Looking at the trend of the lensing signal as function of the color offset, they selected
the sample where the dilution were minimized, obtaining a background sample of $\sim 34$ gal arcmin$^{-2}$ for the computation
of tangential shear profile of the cluster.
The fit of this profile by a  NFW model did not give an acceptable fit: the virial mass computed assuming a NFW profile was $M_{\rm vir}=3.62^{+1.15}_{-0.86}\times h^{-1} 10^{14}M_\odot$ with a  high concentration parameter $c_{\rm vir}=8.87^{+5.22}_{-3.05}$. The same authors also derived the projected mass, obtaining $M_{2D} = 8.69 \times h^{-1} 10^{14}M_\odot$ at the virial radius.

Scaling such mass values to  $h = 0.7$, we derive $M_{\rm vir} \sim (4-5)\times 10^{14}$ $M_\odot$. This value is still consistent
within uncertainties with the value derived in the present analysis 
($M_{\rm vir} = (7.5^{2.7}_{-1.9} \times 10^{14}$ $M_\odot$, $c_{\rm vir} = 5.7^{2.1}_{-1.6}$, case {\em d}).
In our case, we obtain however a better agreement with both the values of $M_{\rm vir}$ and $c_{\rm vir}$ given by the X-ray data,
and a better consistency between parametric and non-parametric  mass estimates, once the projection factor is taken into account.
This may be due either to how the selection of background/foreground galaxies was done, or to a higher accuracy in the shape measurement 
as a combination of the deeper image and calibration of the bias due to SNR (Sect.~\ref{sec:ksbex}).

\begin{figure}
\centering
\includegraphics[width=8 cm]{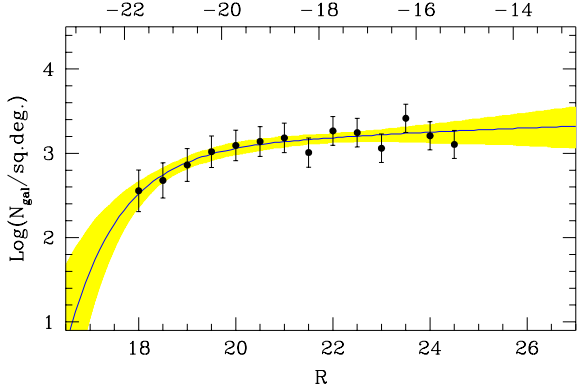}
\caption{$R$-band LF of the galaxies in Abell 383.
Data points are derived by binning the data in magnitude bins of 0.5 mag and error bars
by Poissonian errors;
the curve traces the best-fitting LF and the shaded area marks the model uncertainty,
obtained by a bootstrap technique.}
\label{fig:LF}
\end{figure}

%==================================================================

Finally, we computed the luminosity function (LF hereafter) in the $R$-band
and derived the total luminosity by integrating the fitted Schechter \citep{schechter}
function as described in \citet{Radovich08}. 
In order to obtain the cluster LF, that is the number of galaxies per unit luminosity
and volume belonging to the cluster, we need to remove from our catalog all the
background and foreground galaxies. Usually, this is done by statistically
subtracting the galaxy counts in a control field from galaxy counts in the
cluster direction. Here we take advantage of the selection on the color-color diagrams
described  in Sect.~\ref{sec:results}, and extract a catalog that includes cluster,
foreground and  residual background galaxies.
The last two components were further removed by the statistical subtraction, where we defined as cluster area the circular region
around the cluster center of radius r=10.2 arcmin (1.3Mpc); the control field was instead
defined as the area outside the circle of radius r=15.3 arcmin.
For the best-fitting procedure to the Schechter function, we adopted conventional
routines of minimization on the binned distributions.
Best-fitting parameters are listed in Table~\ref{tab:LF}.
The $R$-band total luminosity, calculated as the Schechter integral, is
$L_{\rm tot}=(2.14\pm0.5)$ $\times$ $10^{12}~L_\odot$.  The errors were estimated by the
propagation of the $68\%$-confidence-errors of each parameter.

For comparison, we then used the relation in \citet{AA.464.451P} between $M_{\rm 200}$ and the optical luminosity, $L_{\rm op}$, 
to see whether the mass  obtained in this paper
is consistent with the value expected for that luminosity. According to this relation, the mass
expected for such luminosity is $M_{\rm 200} = (4.73\pm1.3) \times$ $10^{14}~M_{\odot}$, in
good agreement with the value derived by our weak lensing analysis,
$M_{\rm 200} \sim 6.3 \times$ $10^{14}~M_{\odot}$ (case {\em d}), corresponding to 
$M/L \sim 300 M_\odot/L_\odot$.

\section{Conclusions}

We have computed the cluster Abell 383  mass by weak lensing, using a deep $R$-band image taken with the Suprime camera on the Subaru telescope. Catalogs extracted from combined CFHT+SUBARU $uBVRIz$ images were used to derive photometric redshifts, and improve the weak lensing analysis.
The data were reduced using a pipeline developed in--house, that was specifically designed for wide-field imaging data. The ellipticities, from which the shear signal was derived, were measured using a pipeline based on the KSB approach. We discussed some aspects that may improve the results, namely the size of the window
used to suppress the noise from the outer part of the galaxies, the selection of a limit on SNR below which the measurement ellipticity is not accurate enough, and
a weighting scheme where uncertainties on spatial fitting of the PSF correction terms were taken into account.

The accuracy on the mass estimate by weak lensing available with our KSB pipeline was first derived on simulated images which were built in such a way to
mimic as closely as possible the background noise and the depth of the real image. From these simulations we conclude that the mass can be measured with
an uncertainty $\sim$ 5-10\% for $\log M/M_\odot \geq 14.5$. Such accuracy takes into account the measurement errors of the ellipticity, but not the errors
due to e.g. the  foreground/background galaxy separation, that may introduce an underestimate of the mass.
The impact of such selection was evaluated by comparing three methods for the foreground/background galaxy separation, namely: magnitude cut in one band, color selection and usage of photometric redshifts. All methods gave consistent estimates of the total virial mass, but from the shear profile it can be seen that a dilution of the signal in the inner regions is still present, in the case of a simple magnitude cut. Color selection and photometric redshifts
 provide better results, even if the accuracy of the photometric redshifts is not high due to the few available bands. The virial  mass of Abell 383 here obtained by NFW model fitting is in agreement with the value obtained from the non-parametric mass estimate, that is $M_{\rm vir} \sim 7 \times 10^{14}$ $M_\odot$. Other previous weak lensing analyses give $M_{\rm vir} \sim (4-5) \times 10^{14}$ $M_\odot$: the value found in this paper seems more in agreement with the value found by X-ray data, and we also have a better agreement between parametric and non-parametric estimates, compared to e.g. \cite{Okabe10}.

Finally, we estimated the $R$-band  LF of Abell 383, and derived the total $R$-band luminosity of the cluster: starting from this value and using the relation between mass and luminosity found for clusters by \citet{AA.464.451P}, we conclude that the mass derived by weak lensing is consistent with the value expected for this luminosity.

\section {Acknowledgments}

L.F., Z.H. and M.R. acknowledge the support of the European
Commission Programme 6-th framework, Marie Curie Training and Research
Network ``DUEL'', contract number MRTN-CT-2006-036133.
L.F. was partly supported by the Chinese National Science Foundation
Nos. 10878003 \& 10778725, 973 Program No. 2007CB 815402, Shanghai
Science Foundations and Leading Academic Discipline Project of
Shanghai Normal University (DZL805), and Chen Guang
project with No. 10CG46 of Shanghai Municipal Education
Commission and Shanghai Education Development Foundation.
A.R. acknowledges support from the Italian Space Agency (ASI) contract Euclid-IC I/031/10/0.
We are grateful to the referee for the useful comments that improved this paper.

%\section{}
%\label{sec:}

%\appendix
%\appendixname{VST--Tube}

\bibliography{a383}

\begin{thebibliography}{42}
\expandafter\ifx\csname natexlab\endcsname\relax\def\natexlab#1{#1}\fi

\bibitem[{{Abell} {et~al.}(1989){Abell}, {Corwin}, \& {Olowin}}]{Abell89}
{Abell}, G.~O., {Corwin}, Jr., H.~G., \& {Olowin}, R.~P. 1989, \apjs, 70, 1

\bibitem[{{Bardeau} {et~al.}(2005){Bardeau}, {Kneib}, {Czoske}, {Soucail},
  {Smail}, {Ebeling}, \& {Smith}}]{Bardeau05}
{Bardeau}, S., {Kneib}, J., {Czoske}, O., {et~al.} 2005, \aap, 434, 433

\bibitem[{{Bardeau} {et~al.}(2007){Bardeau}, {Soucail}, {Kneib}, {Czoske},
  {Ebeling}, {Hudelot}, {Smail}, \& {Smith}}]{Bardeau07}
{Bardeau}, S., {Soucail}, G., {Kneib}, J., {et~al.} 2007, \aap, 470, 449

\bibitem[{Bartelmann(1996)}]{Bartelmann96}
Bartelmann, M. 1996, \aap, 313, 697

\bibitem[{{Broadhurst} {et~al.}(2005){Broadhurst}, {Takada}, {Umetsu}, {Kong},
  {Arimoto}, {Chiba}, \& {Futamase}}]{Broadh05}
{Broadhurst}, T., {Takada}, M., {Umetsu}, K., {et~al.} 2005, \apjl, 619, L143

\bibitem[{Bullock {et~al.}(2001)Bullock, Kolatt, Sigad, Somerville, Kravtsov,
  Klypin, Primack, \& Dekel}]{Bullock01}
Bullock, J.~S., Kolatt, T.~S., Sigad, Y., {et~al.} 2001, \mnras, 321, 559

\bibitem[{{Capaccioli} {et~al.}(2005){Capaccioli}, {Mancini}, \&
  {Sedmak}}]{Capaccioli}
{Capaccioli}, M., {Mancini}, D., \& {Sedmak}, G. 2005, The Messenger, 120, 10

\bibitem[{{Capak} {et~al.}(2007){Capak}, {Aussel}, {Ajiki}, {McCracken},
  {Mobasher}, {Scoville}, {Shopbell}, {Taniguchi}, {Thompson}, {Tribiano},
  {Sasaki}, {Blain}, {Brusa}, {Carilli}, {Comastri}, {Carollo}, {Cassata},
  {Colbert}, {Ellis}, {Elvis}, {Giavalisco}, {Green}, {Guzzo}, {Hasinger},
  {Ilbert}, {Impey}, {Jahnke}, {Kartaltepe}, {Kneib}, {Koda}, {Koekemoer},
  {Komiyama}, {Leauthaud}, {Le Fevre}, {Lilly}, {Liu}, {Massey}, {Miyazaki},
  {Murayama}, {Nagao}, {Peacock}, {Pickles}, {Porciani}, {Renzini}, {Rhodes},
  {Rich}, {Salvato}, {Sanders}, {Scarlata}, {Schiminovich}, {Schinnerer},
  {Scodeggio}, {Sheth}, {Shioya}, {Tasca}, {Taylor}, {Yan}, \&
  {Zamorani}}]{capak07}
{Capak}, P., {Aussel}, H., {Ajiki}, M., {et~al.} 2007, \apjs, 172, 99

\bibitem[{Clowe {et~al.}(1998)Clowe, Luppino, Kaiser, Henry, \&
  Gioia}]{apj...497l..61c}
Clowe, D., Luppino, G.~A., Kaiser, N., Henry, J.~P., \& Gioia, I.~M. 1998,
  \apj, 497L, 61

\bibitem[{{Ebeling} {et~al.}(1996){Ebeling}, {Voges}, {Bohringer}, {Edge},
  {Huchra}, \& {Briel}}]{ebeling}
{Ebeling}, H., {Voges}, W., {Bohringer}, H., {et~al.} 1996, \mnras, 281, 799

\bibitem[{Erben {et~al.}(2001)Erben, Waerbeke, Bertin, Mellier, \&
  Schneider}]{erben01}
Erben, T., Waerbeke, L.~V., Bertin, E., Mellier, Y., \& Schneider, P. 2001,
  \aap, 366, 717

\bibitem[{Fahlman {et~al.}(1994)Fahlman, Kaiser, Squires, \& Woods}]{fahlman}
Fahlman, G., Kaiser, N., Squires, G., \& Woods, D. 1994, \apj, 437, 56

\bibitem[{{Feldmann} {et~al.}(2006){Feldmann}, {Carollo}, {Porciani}, {Lilly},
  {Capak}, {Taniguchi}, {Le F{\`e}vre}, {Renzini}, {Scoville}, {Ajiki},
  {Aussel}, {Contini}, {McCracken}, {Mobasher}, {Murayama}, {Sanders},
  {Sasaki}, {Scarlata}, {Scodeggio}, {Shioya}, {Silverman}, {Takahashi},
  {Thompson}, \& {Zamorani}}]{Feldmann06}
{Feldmann}, R., {Carollo}, C.~M., {Porciani}, C., {et~al.} 2006, \mnras, 372,
  565

\bibitem[{{Fetisova} {et~al.}(1993){Fetisova}, {Kuznetsov}, {Lipovetskij},
  {Starobinskij}, \& {Olowin}}]{Fetisova93}
{Fetisova}, T.~S., {Kuznetsov}, D.~Y., {Lipovetskij}, V.~A., {Starobinskij},
  A.~A., \& {Olowin}, R.~P. 1993, Astronomy Letters, 19, 198

\bibitem[{{Gill} {et~al.}(2009){Gill}, {Young}, {Draskovic}, {Honscheid},
  {Lin}, {Kuropatkin}, {Martini}, {Peeples}, {Rozo}, {Smith}, \&
  {Weinberg}}]{Gill09}
{Gill}, M.~S.~S., {Young}, J.~C., {Draskovic}, J.~P., {et~al.} 2009,
  ArXiv:0909.3856

\bibitem[{Grado {et~al.}(2011)Grado, Capaccioli, Limatola, \& Getman}]{Grado10}
Grado, A., Capaccioli, M., Limatola, L., \& Getman, F. 2011, ArXiv:1102.1588

\bibitem[{{Heymans} {et~al.}(2006){Heymans}, {Van Waerbeke}, {Bacon}, {Berge},
  {Bernstein}, {Bertin}, {Bridle}, {Brown}, {Clowe}, {Dahle}, {Erben}, {Gray},
  {Hetterscheidt}, {Hoekstra}, {Hudelot}, {Jarvis}, {Kuijken}, {Margoniner},
  {Massey}, {Mellier}, {Nakajima}, {Refregier}, {Rhodes}, {Schrabback}, \&
  {Wittman}}]{step1}
{Heymans}, C., {Van Waerbeke}, L., {Bacon}, D., {et~al.} 2006, \mnras, 368,
  1323

\bibitem[{{Hoekstra}(2003)}]{Hoekstra03}
{Hoekstra}, H. 2003, \mnras, 339, 1155

\bibitem[{Hoekstra(2007)}]{hoekstra07}
Hoekstra, H. 2007, \mnras, 379, 317

\bibitem[{Hoekstra {et~al.}(1998)Hoekstra, Franx, Kuijken, \&
  Squires}]{Hoekstra98}
Hoekstra, H., Franx, M., Kuijken, K., \& Squires, G. 1998, \apj, 504, 636

\bibitem[{{Hoekstra} {et~al.}(2010){Hoekstra}, {Hartlap}, {Hilbert}, \& {van
  Uitert}}]{Hoekstra10}
{Hoekstra}, H., {Hartlap}, J., {Hilbert}, S., \& {van Uitert}, E. 2010,
  ArXiv:1011.1084

\bibitem[{Luppino \& Kaiser(1997)}]{kaiser97}
Luppino, G. \& Kaiser, N. 1997, \apj, 475, 20

\bibitem[{{Massey} {et~al.}(2007){Massey}, {Heymans}, {Berg{\'e}}, {Bernstein},
  {Bridle}, {Clowe}, {Dahle}, {Ellis}, {Erben}, {Hetterscheidt}, {High},
  {Hirata}, {Hoekstra}, {Hudelot}, {Jarvis}, {Johnston}, {Kuijken},
  {Margoniner}, {Mandelbaum}, {Mellier}, {Nakajima}, {Paulin-Henriksson},
  {Peeples}, {Roat}, {Refregier}, {Rhodes}, {Schrabback}, {Schirmer}, {Seljak},
  {Semboloni}, \& {van Waerbeke}}]{step2}
{Massey}, R., {Heymans}, C., {Berg{\'e}}, J., {et~al.} 2007, \mnras, 376, 13

\bibitem[{{Medezinski} {et~al.}(2010){Medezinski}, {Broadhurst}, {Umetsu},
  {Oguri}, {Rephaeli}, \& {Ben{\'{\i}}tez}}]{Medez10}
{Medezinski}, E., {Broadhurst}, T., {Umetsu}, K., {et~al.} 2010, \mnras, 405,
  257

\bibitem[{{Miyazaki} {et~al.}(2002){Miyazaki}, {Komiyama}, {Sekiguchi},
  {Okamura}, {Doi}, {Furusawa}, {Hamabe}, {Imi}, {Kimura}, {Nakata}, {Okada},
  {Ouchi}, {Shimasaku}, {Yagi}, \& {Yasuda}}]{suprimecam}
{Miyazaki}, S., {Komiyama}, Y., {Sekiguchi}, M., {et~al.} 2002, \pasj, 54, 833

\bibitem[{{Oguri} {et~al.}(2010){Oguri}, {Takada}, {Okabe}, \&
  {Smith}}]{Oguri10}
{Oguri}, M., {Takada}, M., {Okabe}, N., \& {Smith}, G.~P. 2010, \mnras, 405,
  2215

\bibitem[{{Okabe} {et~al.}(2010){Okabe}, {Takada}, {Umetsu}, {Futamase}, \&
  {Smith}}]{Okabe10}
{Okabe}, N., {Takada}, M., {Umetsu}, K., {Futamase}, T., \& {Smith}, G.~P.
  2010, \pasj, 62, 811

\bibitem[{{Pickles}(1985)}]{Pickles}
{Pickles}, A.~J. 1985, \apjs, 59, 33

\bibitem[{Popesso {et~al.}(2007)Popesso, Biviano, B\"oringher, \&
  Romaniello}]{AA.464.451P}
Popesso, P., Biviano, A., B\"oringher, H., \& Romaniello, M. 2007, \aap, 464,
  451

\bibitem[{{R Development Core Team}(2010)}]{R}
{R Development Core Team}. 2010, R: A Language and Environment for Statistical
  Computing, R Foundation for Statistical Computing, Vienna, Austria, {ISBN}
  3-900051-07-0

\bibitem[{{Radovich} {et~al.}(2008){Radovich}, {Puddu}, {Romano}, {Grado}, \&
  {Getman}}]{Radovich08}
{Radovich}, M., {Puddu}, E., {Romano}, A., {Grado}, A., \& {Getman}, F. 2008,
  \aap, 487, 55

\bibitem[{{Rizza} {et~al.}(1998){Rizza}, {Burns}, {Ledlow}, {Owen}, {Voges}, \&
  {Bliton}}]{Rizza98}
{Rizza}, E., {Burns}, J.~O., {Ledlow}, M.~J., {et~al.} 1998, \mnras, 301, 328

\bibitem[{{Romano} {et~al.}(2010){Romano}, {Fu}, {Giordano}, {Maoli},
  {Martini}, {Radovich}, {Scaramella}, {Antonuccio-Delogu}, {Donnarumma},
  {Ettori}, {Kuijken}, {Meneghetti}, {Moscardini}, {Paulin-Henriksson},
  {Giallongo}, {Ragazzoni}, {Baruffolo}, {Dipaola}, {Diolaiti}, {Farinato},
  {Fontana}, {Gallozzi}, {Grazian}, {Hill}, {Pedichini}, {Speziali},
  {Smareglia}, \& {Testa}}]{Romano10}
{Romano}, A., {Fu}, L., {Giordano}, F., {et~al.} 2010, \aap, 514, A88+

\bibitem[{{Schechter}(1976)}]{schechter}
{Schechter}, P. 1976, \apj, 203, 297

\bibitem[{Schirmer {et~al.}(2004)Schirmer, Erben, Schneider, Wolf, \&
  Meisenheimer}]{aa...420...75s}
Schirmer, M., Erben, T., Schneider, P., Wolf, C., \& Meisenheimer, K. 2004,
  \aap, 420, 75

\bibitem[{{Schlegel} {et~al.}(1998){Schlegel}, {Finkbeiner}, \&
  {Davis}}]{schlegel}
{Schlegel}, D.~J., {Finkbeiner}, D.~P., \& {Davis}, M. 1998, \apj, 500, 525

\bibitem[{{Schmidt} \& {Allen}(2007)}]{Schmidt07}
{Schmidt}, R.~W. \& {Allen}, S.~W. 2007, \mnras, 379, 209

\bibitem[{Schrabback {et~al.}(2010)Schrabback, Hartlap, Joachimi, Kilbinger,
  Simon, Benabed, Brada\v{c}, Eifler, Erben, \& Fassnacht}]{schrabback10}
Schrabback, T., Hartlap, J., Joachimi, B., {et~al.} 2010, \aap, 516, 63

\bibitem[{{Smith} {et~al.}(2001){Smith}, {Kneib}, {Ebeling}, {Czoske}, \&
  {Smail}}]{smith01}
{Smith}, G.~P., {Kneib}, J., {Ebeling}, H., {Czoske}, O., \& {Smail}, I. 2001,
  \apj, 552, 493

\bibitem[{Venables \& Ripley(2002)}]{venables02}
Venables, W.~N. \& Ripley, B.~D. 2002, Modern Applied Statistics with S, 4th
  edn. (New York: Springer), isbn 0-387-95457-0

\bibitem[{{Voges}(1992)}]{Voges92}
{Voges}, W. 1992, in European Southern Observatory Conference and Workshop
  Proceedings, Vol.~43, European Southern Observatory Conference and Workshop
  Proceedings, ed. {A.~Heck \& F.~Murtagh}, 139--+

\bibitem[{Wright \& Brainerd(2000)}]{Wright00}
Wright, C.~O. \& Brainerd, T.~G. 2000, \apj, 534, 34

\end{thebibliography}

\end{document}